\documentclass[hyper,a4paper]{SIMO}  
\usepackage{graphicx}
\usepackage{epsfig,rotating}
\usepackage{exscale}
\usepackage{amsmath}
\usepackage{cite}
\usepackage{latexsym}
\usepackage{graphics}
\usepackage[english]{babel}

\newcommand{\ve}{\varepsilon}

\newcommand{\bm}{\begin{minipage}}
\renewcommand{\em}{\end{minipage}}
\newcommand{\bi}{\begin{itemize}}
\newcommand{\ei}{\end{itemize}}
\newcommand{\be}{\begin{eqnarray}}
\newcommand{\en}{\end{eqnarray}}
\newcommand{\ba}{\begin{array}}
\newcommand{\ea}{\end{array}}
\newcommand{\bc}{\begin{center}}
\newcommand{\ec}{\end{center}}
\newcommand{\bfi}{\begin{figure}}
\newcommand{\efi}{\end{figure}}
\newcommand{\im}{\mbox{Im}}
\newcommand{\re}{\mbox{Re}}
\newcommand{\bfr}{\begin{flushright}}
\newcommand{\efr}{\end{flushright}}

\newcommand{\no}{\nonumber}
\newcommand{\ra}{\rightarrow}

\newcommand{\ov}[1]{\overline{#1}}

\newcommand{\f}{f_0}

\title{
A practical guide\\ to unravel time-like transition form factors 
}
\author{Simone Pacetti\\
Enrico Fermi Center, Rome, Italy\\
INFN, Laboratori Nazionali di Frascati, Frascati, Italy  \\ 
E-mail: 
\email{simone.pacetti@lnf.infn.it} }
\abstract{%
A method to determine masses, widths and coupling constants of vector mesons, 
like $\phi(1020)$, $\omega(782)$ and $\rho^0(770)$ recurrences is defined. 
Starting from data on decay rates and cross sections for the processes:
$\phi\ra M_I\gamma$, $\phi\ra M_I e^+e^-$ and $e^+e^-\ra M_I\phi$, 
where $M_I$ is a pseudoscalar or scalar meson with isospin $I=0,1$,
the time-like transition form factors, which describe the vertex $\phi\gamma M_I$, 
are parametrized using a vector meson-propagators description in 
the low energy region ($<3-4$ GeV), the quark-counting rule prescription for
the high energy behavior, and the analyticity imposed by means of the
dispersion relations.
}
\begin{document}
%
%
%
%
%
\section{Introduction}
\label{intro}
Recently the interest in the electromagnetic hadron structures has 
increased considerably due to the large amount of data coming from 
both flavor factories~\cite{flavor}, working with initial state 
radiation (ISR), and fixed-target machines operating with polarized 
beams\cite{jlab}.\\
The study of phenomenological properties of hadrons: the magnetic moments,
structure functions, number, species and mixing of quarks and gluons
in their wave functions, etc. plays a fundamental role in our
understanding of the QCD dynamics. The analysis of electromagnetic
decays and hadron-photon interactions provides unique information for 
this purpose. Indeed, since the photon-hadron coupling occurs through the 
direct interaction of the photon with the electric charges of single quarks,
such processes represent probes by the intrinsic structure of hadrons.
\subsection[``Static'' form factors]
{``Static'' form factors~\cite{landsberg}}
\label{static-ff}
The first attempt to unravel the electromagnetic structure of a hadron
by studying a form factor (ff) was done by Rutherford in his famous
experiment. In that case he studied the scattering of a charged probe, 
the $^4He$ nucleus, in the electromagnetic field of atoms of gold.
The intensity of the scattered $^4He$ beam as a function of the scattering
angle was just a measure of the electromagnetic ff as a function of the 
transferred momentum.
\bfi[h!]
\bc
\epsfig{file=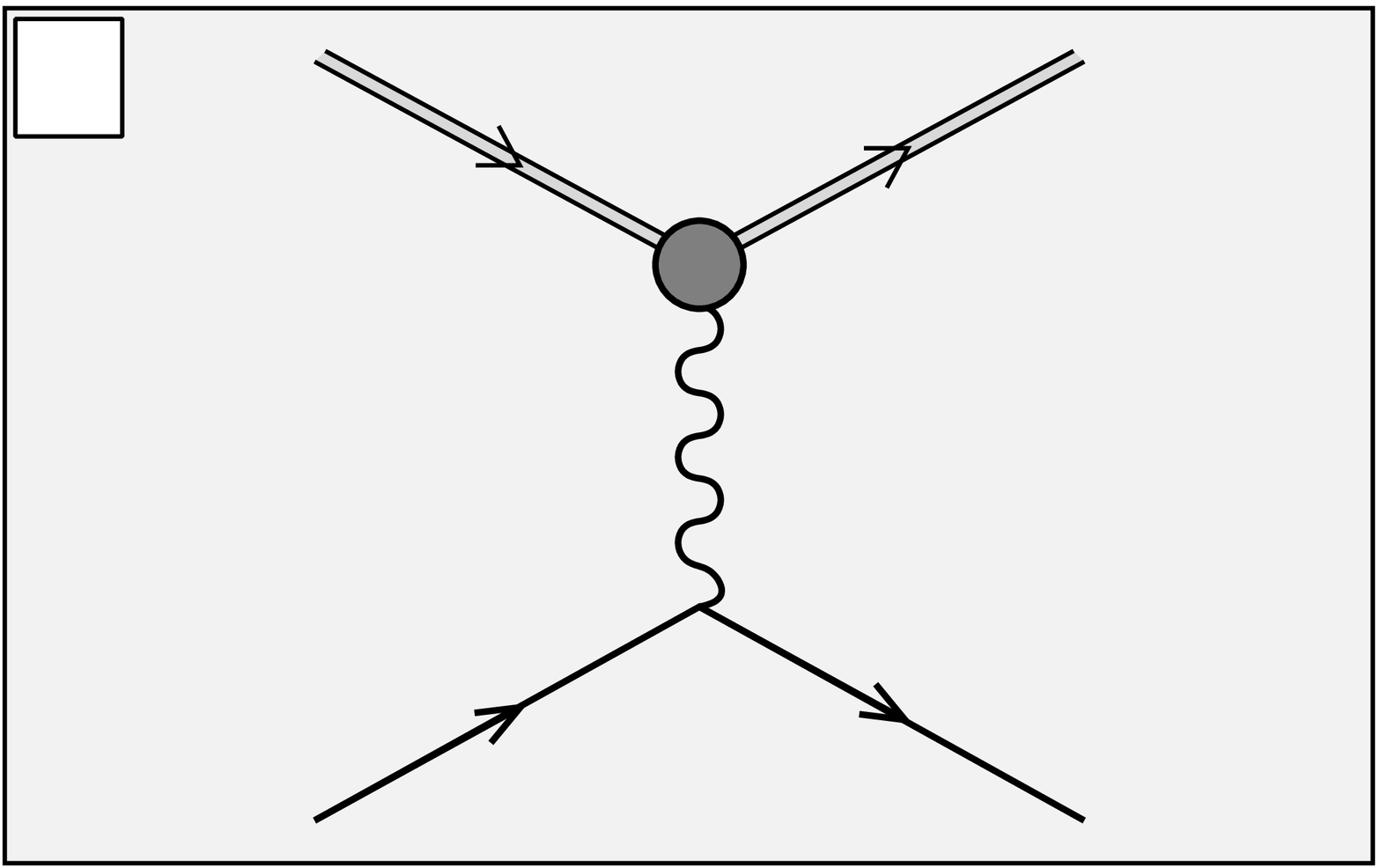,height=30mm}
\put(-114,16){$e^-\!(\!k_1\!)$}
\put(-46,16){$e^-\!(\!k_2)$}
\put(-114,60){$\pi^+\!(\!p_1\!)$}
\put(-46,60){$\pi^+\!(\!p_2\!)$}
\put(-90,40){$\gamma(q)$}
\put(-51,40){$q^2<0$}
\put(-80,70){$F_\pi\!(\!q^2\!)$}
\put(-132,74){$a$}
\hspace{20mm}
\epsfig{file=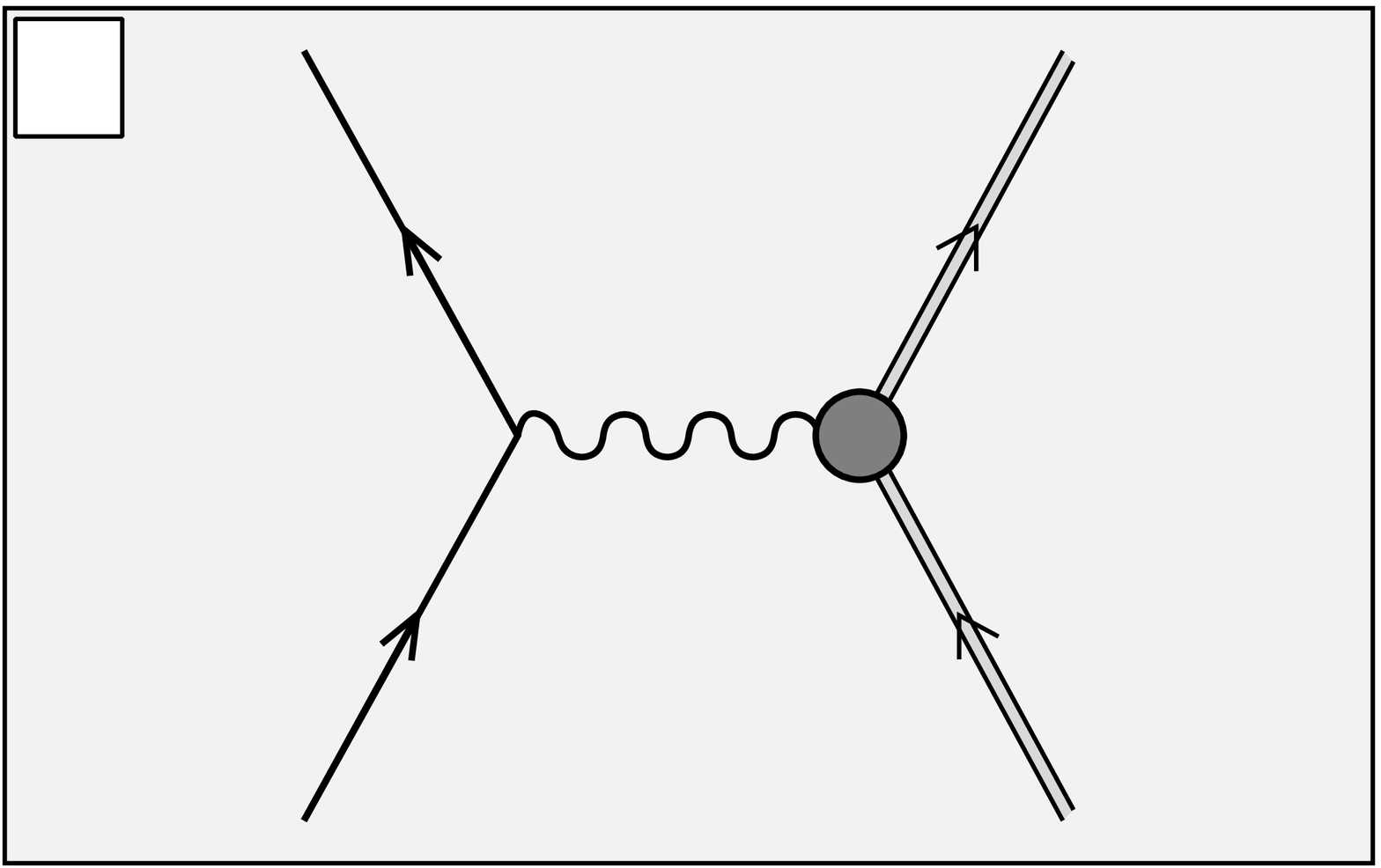,height=30mm}
\put(-128,60){$e^+\!(\!\mbox{-}k_2\!)$}
\put(-124,18){$e^-\!(\!k_1\!)$}
\put(-36,60){$\pi^+\!(\!p_1\!)$}
\put(-36,18){$\pi^-\!(\!\mbox{-}p_2\!)$}
\put(-44,39.5){$F_\pi\!(\!q^2\!)$}
\put(-79,50){$\gamma(q)$}
\put(-84,20){$q^2>0$}
\put(-132,74){$b$}
\caption{\label{e-pi}
One-photon exchange diagrams: for the scattering $e^-\pi^+\ra e^-\pi^+$ ($a$) 
and for the annihilation $e^+e^-\ra \pi^+\pi^-$ ($b$).
}
\ec
\efi\\
We consider in more details the photon-pion interaction, i.e. the vertex 
$\pi\gamma\pi$ shown in fig.~\ref{e-pi}, in case of scattering 
$e^-\pi^+\ra e^-\pi^+$ (fig.~\ref{e-pi}$a$) and annihilation 
$e^+ e^-\ra \pi^+\pi^-$ (fig.~\ref{e-pi}$b$).
\\
For pointlike particles, as the leptons, the electromagnetic
current which describes the vertex $e\gamma e$ (fig.~\ref{e-pi}$a,b$)
has the form $J_e^\mu=-ie\ov{u}(k_2)\gamma^\mu u(k_1)$ (with 4-momenta as 
labeled in fig.~\ref{e-pi}).
The electromagnetic current, in the case of the hadronic vertex $\pi\gamma\pi$, 
is
\be
J_\pi^\mu=\underbrace{ie(p_1+p_2)^\mu}_{\rm pointlike} F_\pi(q^2),
\label{j-pi}
\en
where $p_1$ and $p_2$ are the 4-momenta (see fig.~\ref{e-pi}), and 
$F_\pi(q^2)$, the pion ff, is a scalar function of the 
transferred momentum squared. The tensor part of this current, called 
``pointlike'' in expression~(\ref{j-pi}), is achieved by requiring that
the most general tensorial form, which derives from the spin properties of
the particles under consideration, fulfills Lorentz and gauge 
invariance~\footnotemark[1]. %
\footnotetext[1]{
Charge conjugation invariance requires a vanishing neutral pion ff.
}%
The ff gives an exhaustive characterization
of the spatial distribution of charge of the pion, i.e. it accounts
for the extended hadronic structure of the particle. In the nonrelativistic 
limit, the ff tends to the Fourier transformation of the charge
density distribution.\\
For high-energy photon-hadron interactions the relation 
with the charge density becomes more complex than the simple Fourier 
transformation. In spite of that, the ff describes unambiguously 
the electromagnetic structure of the particle and it represents 
the only quantity, intrinsically related to this structure, which is 
experimentally accessible. 
\\
Using the previous expressions for the leptonic and hadronic currents,
the differential cross section for the scattering $e^-\pi^+\ra e^-\pi^+$,
in Born approximation (see fig.~\ref{e-pi}), can be written as:
\be
\frac{d\sigma(e^-\pi^+\ra e^-\pi^+)}{dq^2}=\left[\frac{d\sigma
(e^-\pi^+\ra e^-\pi^+)}{dq^2}\right]_{\rm pointlike} 
\left[F_\pi(q^2)\right]^2.
\label{pi-sl}
\en
The values of the ff may be extracted by comparing cross section data 
and the differential cross section computed in the case of pointlike charged 
hadrons, $\left[d\sigma/dq^2\right]_{\rm pointlike} $. By using the
scattering cross section data, only space-like values can be
investigated. In this case $q^2=(p_1-p_2)^2\le0$. In the
space-like region the ff is a real function, hence the data on
the differential cross section $\frac{d\sigma(e^-\pi^+\ra e^-\pi^+)}{dq^2}$
represent a direct measurement of the ff itself. 
\\
The annihilation $e^+ e^-\ra \pi^+\pi^-$, shown in fig.~\ref{e-pi}$b$
in case of one-photon exchange, is the scattering crossed process. 
It contains the same vertices and the cross section,
computed by contracting the same currents with opposite signs for
antiparticle 4-momenta, reads:
\be
\sigma(e^+ e^-\ra \pi^+\pi^-)=
[\sigma(e^+ e^-\ra \pi^+\pi^-)]_{\rm pointlike}\left|F_\pi(q^2)\right|^2.
\label{pi-tl}
\en
In this case $q^2=(p_1-p_2)^2\ge 4M_\pi^2$, so by measuring the annihilation
cross section we are probing the ff in the portion of the time-like region above
the physical threshold $4M_\pi^2$. Here the ff is a complex function and only
its modulus is experimentally accessible [see expression~(\ref{pi-tl})].
No information about the complex structure of this function, i.e. real and 
imaginary parts, may be obtained by direct experimental observation.
\bfi[h!]
\bc
\epsfig{file=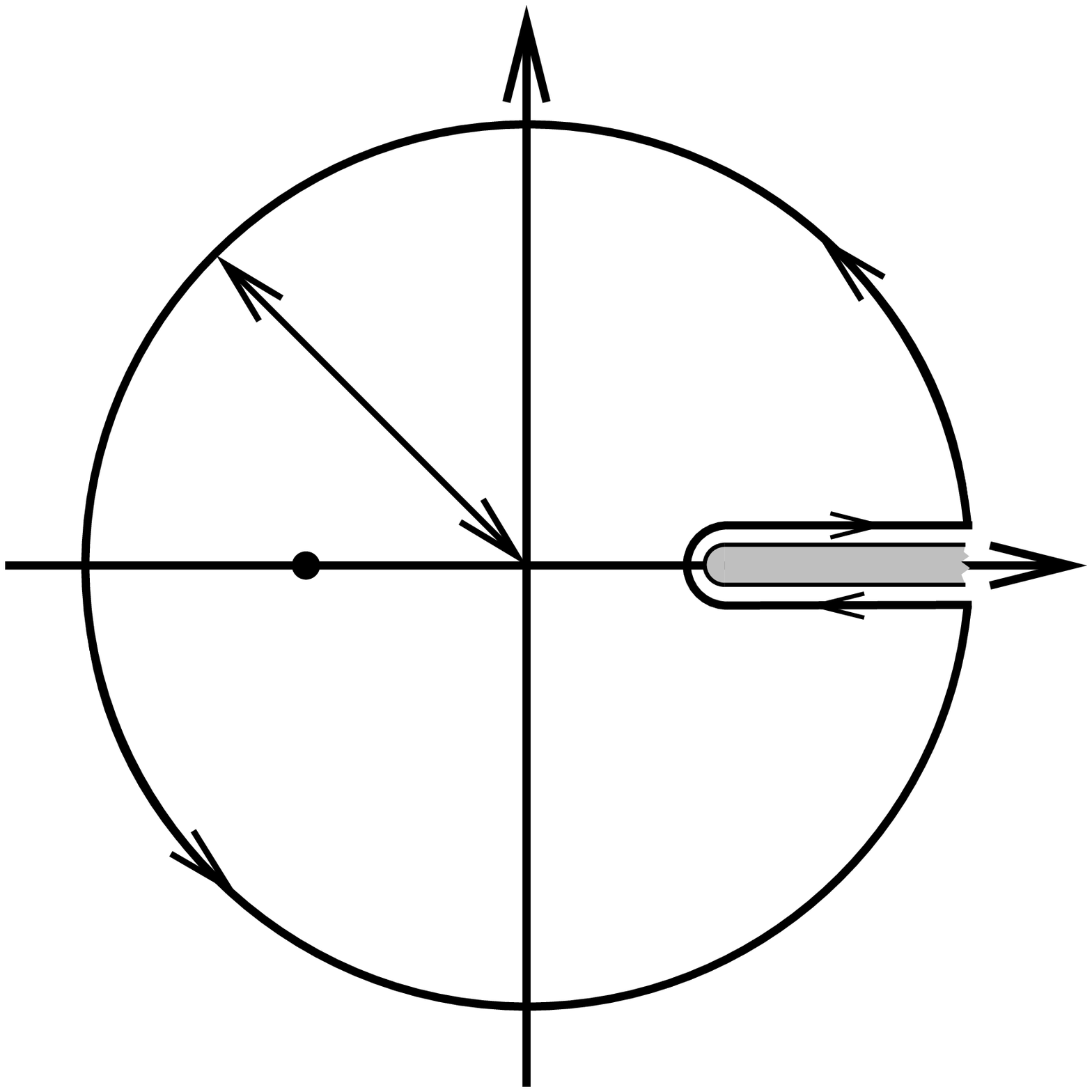,width=60mm}
\put(-84,160){$\im(q^2)$}
\put(-15,68){$\re(q^2)$}
\put(-65,70){$s_{\rm th}$}
\put(-125,71){$q^2$}
\put(-112,108){$\mathcal{R}$}
\put(-35,133){$\mathcal{C}$}
\caption{\label{cauchy}
Analyticity domain for the pion ff. The shadowed band
indicates the cut over the real axis. The space-like 
$q^2$, on the left, is a generic point where the real 
values of the ff may be extracted from the scattering 
cross section.
}
\ec
\efi\\%
Nevertheless, by invoking unitarity and analyticity for the ff's, 
integral relations between the modulus and the phase above threshold may be 
established. By means of these relations, the phase is computed as a function 
of the modulus and therefore we gain a complete knowledge of the complex 
structure of the ff's.
\\%
In general (see fig.~\ref{cauchy}), the ff's are analytic functions in the 
$q^2$-complex plane with a cut, on the real axis, starting from the theoretical 
threshold $s_{\rm th}=4M_\pi^2$, which, in this case, corresponds to the physical 
one, up to infinity.
%
%
We introduce here the concepts of theoretical and physical threshold for
a ff. Both of them lie in the time-like region and refer to the annihilation
process.
\begin{itemize}
\item The theoretical threshold, called $s_{\rm th}$, is the energy of the 
      first hadronic channel that carries the same quantum numbers of the final
      state under consideration, e.g. in this case $\pi^+\pi^-$, produced 
      through the $e^+e^-$ annihilation.\\
\item The physical threshold corresponds to the production energy of the hadronic 
      final state, it is greater or, only in few cases, equal to the physical one.
\end{itemize} 
The pion ff is just one of these special cases where the thresholds correspond 
to the same energy.
The discontinuity over the real axis is the superposition of infinite cuts, 
each of which corresponds to the opening of a channel for every allowed 
final state. In principle these functions are well defined for all complex 
$q^2$ outside the cut, but only real values of $q^2$ are experimentally 
accessible. 
\\%
The case of the pion ff is particularly simple, mainly due to
the fact that it is a boson with spin zero. The study of nucleon
ff's requires a more complicated treatment. These particles are 
fermions and their spin structures are described by two, instead of 
only one, ff's. Nevertheless, the scheme and the method of the analysis 
remain the same.
\bfi[h!]
\epsfig{file=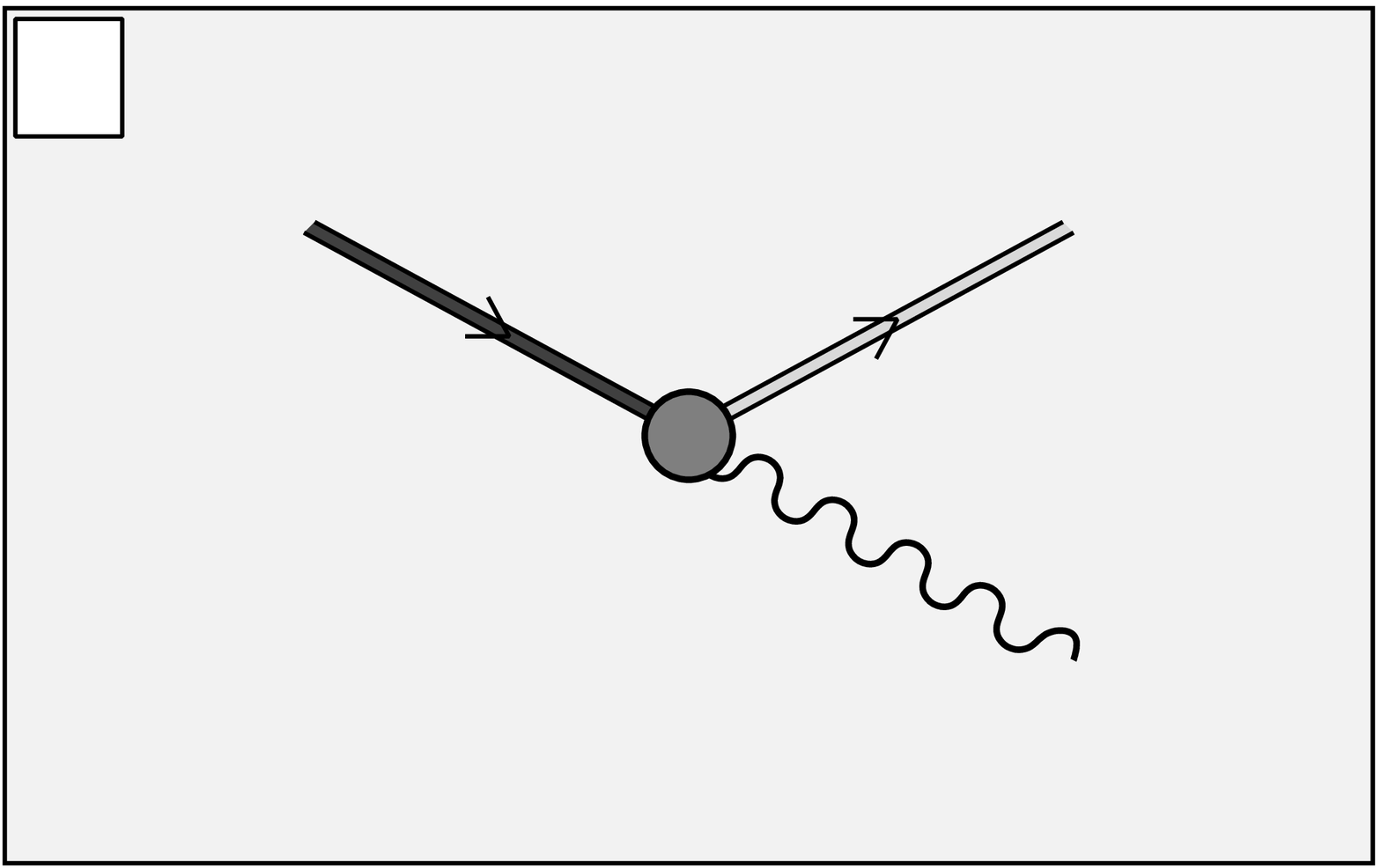,height=30mm}
\put(-115,60){$A$}
\put(-29,60){$B$}
\put(-28,20){$\gamma$}
\put(-84,58){$F_{AB}(0)$}
\put(-85,-15){$q^2=0$}
\put(-132,74){$a$}
\hspace{1mm}
\epsfig{file=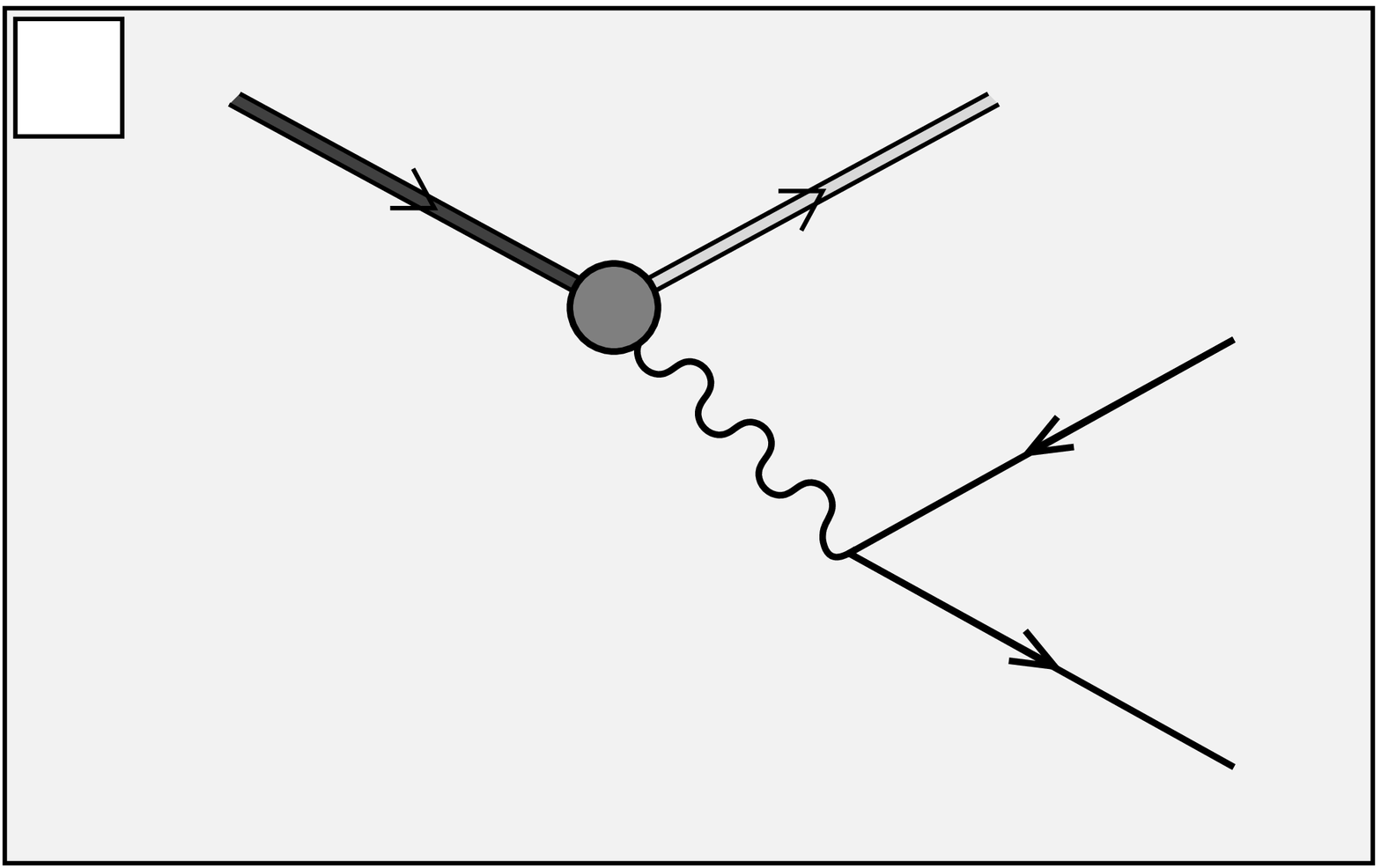,height=30mm}
\put(-122,72){$A$}
\put(-36,72){$B$}
\put(-62,46){$\gamma(q)$}
\put(-95,70){$F_{AB}(q^2)$}
\put(-13,50){$e^+$}
\put(-13,7){$e^-$}
\put(-132,74){$b$}
\put(-128,-15){$(2m_e)^2<q^2\!<\!(\!M_A\!\!-\!\!M_B\!)^2$}
\hspace{1mm}
\epsfig{file=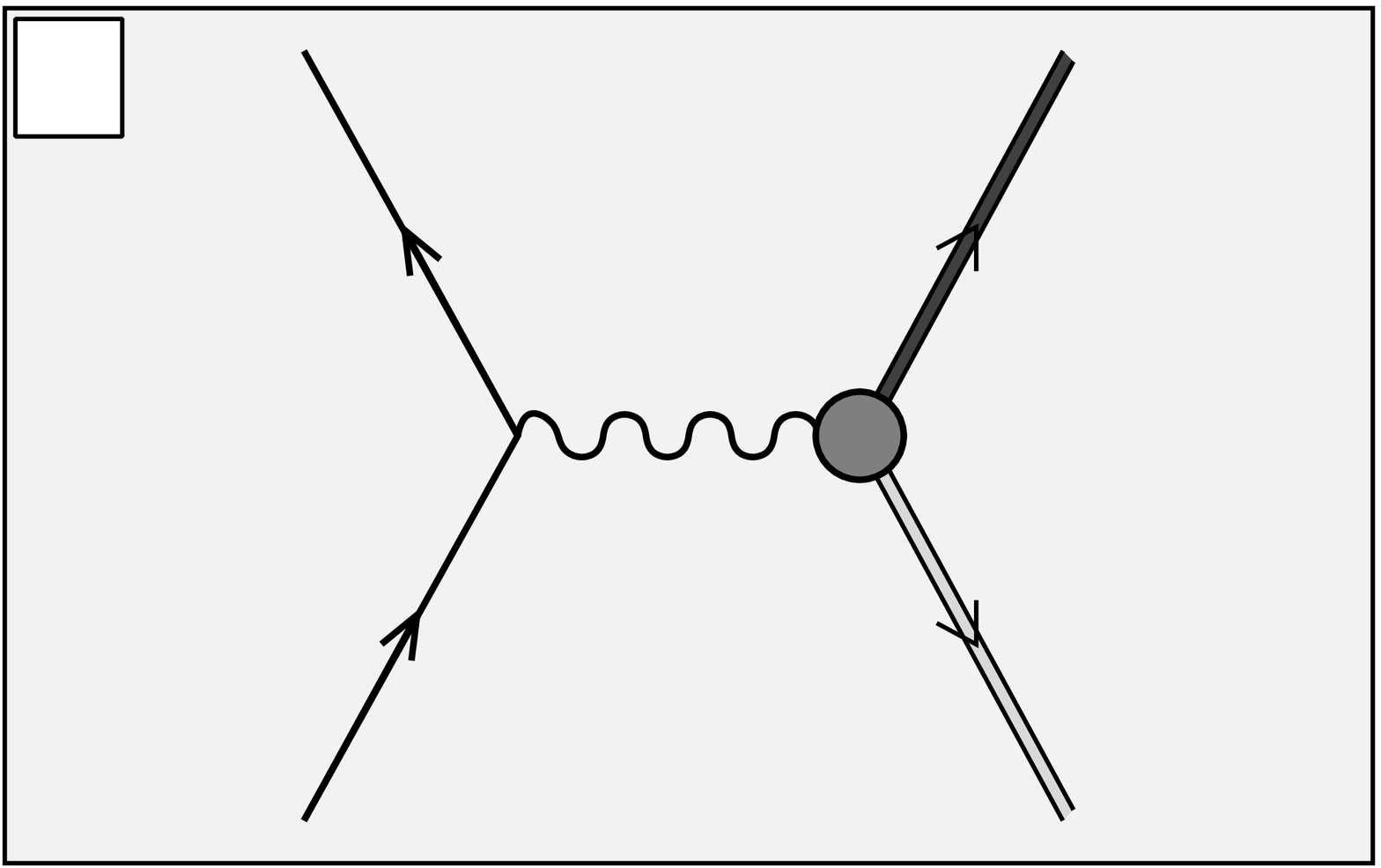,height=30mm}
\put(-120,75){$e^+$}
\put(-120,3){$e^-$}
\put(-28,75){$A$}
\put(-28,2){$B$}
\put(-44,39.5){$F_{AB}(q^2)$}
\put(-79,50){$\gamma(q)$}
\put(-103,-15){$q^2\!>\!(\!M_A\!\!+\!\!M_B\!)^2$}
\put(-132,74){$c$}
\caption{\label{agb}
Diagrams for the radiative decay $A\ra B\gamma$ ($a$),
the conversion decay $A\ra Be^+e^-$ ($b$) and
annihilation $e^+e^-\ra AB$ ($c$). In all these cases
the coupling $A\gamma B$ is described by the same tff $F_{AB}(q^2)$,
but evaluated in the different regions of $q^2$.
}
\efi
\subsection[``Dynamic'' time-like form factors]
{``Dynamic'' time-like form factors~\cite{landsberg}}
\label{dynamic-ff}
As we have already stated the ff's provide a crucial mean to 
investigate the internal hadron structure. They describe the
photon-hadron interaction in terms of coupling between photons
and charges of quark constituents, and hence they give
unique information about the QCD dynamics inside a hadron.  
However, up to now we have dealt with only ``static'' structure 
functions, i.e. ff's that, due to the charge conjugation 
invariance~\footnotemark[2], are defined only for charged mesons.
\footnotetext[2]{%
One-photon exchange processes (see fig.~\ref{e-pi}) for neutral
mesons have vanishing amplitude, and then vanishing ff for all values
of $q^2$, since the charge-conjugation parity is conserved in electromagnetic 
interactions. 
}%
To study the internal structure of neutral hadrons, we
may consider vertices like $A\gamma B$, where a neutral particle
$A$ converts into a different neutral particle $B$ with the emission 
of a photon while conserving all the quantum numbers. The physical
process would be (see fig.~\ref{agb}$a$):
\be
A\ra B+\gamma, \hspace{10mm}\mbox{charge conjugation conservation}
\Longrightarrow
C_A=C_B\cdot C_\gamma,
\label{a-bg}
\en
where, to conserve charge conjugation, being $C_\gamma=-1$, we have: 
$C_A=-C_B$. For every pair of neutral mesons for which the radiative 
decay~(\ref{a-bg}) is allowed, these other one-photon exchange processes
also are allowed:
\bi
\item conversion decay (see fig.~\ref{agb}$b$): 
      $A\ra B \gamma^*\ra Be^+e^-$ with $(2m_e)^2<q^2<(M_A-M_B)^2$ ($m_e$ is the
      electron mass);
\item annihilation (see fig.~\ref{agb}$c$): 
      $e^+e^-\ra \gamma^*\ra AB$ with $q^2\ge (M_A+M_B)^2$.
\ei
The probability for the meson $A$ to convert into a meson $B$ by
emitting a photon of 4-momentum $q$ is described by a specific ff called 
the ``dynamic'' or transition form factor (tff) $F_{AB}(q^2)$. 
Contrary to the ordinary ``static'' ff's which, being related to 
the structure of a single particle, vanish for neutral mesons, the tff's, 
that describe the electromagnetic structure of 
transition vertex, i.e. the dynamics of the conversion of a meson 
into another, are, in principle, different from zero.
To extract values of a tff from experimental data on
decay rates and annihilation cross section we define the dynamic
conversion current:
\be
J_{AB}^\mu=\mathcal{J}^\mu_{\rm pointlike}(q^2)\cdot F_{AB}(q^2),
\label{j-ab}
\en
where the tensor $\mathcal{J}^\mu_{\rm pointlike}(q^2)$, obtained
by requiring Lorentz and gauge invariance, describes
the vertex $A\gamma B$ in case of structureless particles
and the tff $F_{AB}(q^2)$ accounts for the modifications of this
vertex due to the extended electromagnetic internal structures of 
the particles $A$ and $B$. \\
In more detail, to investigate the tff $F_{AB}(q^2)$, three experimental 
observables may be considered:
\bi
\item the radiative decay rate that, by using the current~(\ref{j-ab}),
      has the form:
      \be
      \Gamma(A\ra B\gamma)=\mathcal{G}_{\rm pointlike}\cdot
      \left[F_{AB}(0)\right]^2,
      \label{k-g}
      \en
      this is a constant quantity which depends on the real value
      of the tff at $q^2=0$ (real photon). 
\item The differential decay rate:
\be
\frac{d\Gamma(A\ra Be^+e^-)}{dq^2}=
\left[\frac{d\mathcal{G}}{dq^2}\right]_{\rm pointlike}\cdot
\left|F_{AB}(q^2)\right|^2,
      \label{k-dg}
\en
with: $(2m_e)^2<q^2<(M_A-M_B)^2$. 
\item The annihilation cross section:
\be
\sigma(e^+e^-\ra AB)=\mathcal{S}_{\rm pointlike}\cdot
\left|F_{AB}(q^2)\right|^2,
      \label{k-s}
\en
with: $q^2>(M_A+M_B)^2$. 
\ei
The quantities $\mathcal{G}_{\rm pointlike}$, 
$\left[d\mathcal{G}/dq^2\right]_{\rm pointlike}$, and 
$\mathcal{S}_{\rm pointlike}$ represent the kinematic factors 
obtained using the current~(\ref{j-ab}) with $F_{AB}(q^2)=1$.
As we will see in the following [\textsection\ref{tff-data}, 
eq.~(\ref{dimensions})], an interpretation in terms of pointlike decay 
rates and annihilation cross section of these quantities, obtained by 
using only the pointlike part of the current~(\ref{j-ab}), 
appears awkward because of the wrong dimensions.
\\
Finally we note that, although the tff's, exactly as the ff's,
are analytic functions defined in the whole $q^2$-complex plane with
the cut $(s_{\rm th},\infty)$, the listed processes cover only a portion
of the time-like region. In principle the space-like region could be 
investigated by measuring the ``conversion scattering'' $Ae^-\ra Be^-$, 
however the enormous difficulty in realizing a stable massive meson beam,
makes this possibility a very hard task. In the time-like region there is
an interval which is theoretically not accessible by the experiments,
it is the so-called ``unphysical'' region: $[(M_A-M_B),(M_A+M_B)]$. 
%
%
\\
In the rest of  this work we will focus on a particular group of tff's,
i.e. those describing the $\phi(1020)\gamma M$ conversion, where 
$\phi(1020)$ is $s\ov{s}$ vector meson and $M$ is a generic light 
(made of $u$, $d$, and $s$ quarks)
pseudoscalar or scalar meson.
\section{Strategy}
\label{strategy}
We define a general procedure to study the tff of a generic conversion 
$\phi\gamma M$, where, from now on, $\phi$ stands for the vector meson 
$\phi(1020)$ and $M$ for a light pseudoscalar or scalar meson.
By assuming that in a certain $q^2$ interval, called the resonance region,
the photon couples with the mesons $\phi$ and $M$ through a series of 
intermediate vector mesons, we parametrize the tff in this region 
as a sum of propagators weighted by the corresponding coupling constants. 
This parameterization is then extended, to higher values of $q^2$ using
the power law asymptotic behavior provided by the quark-counting rule (QcR)~\cite{brodsky}, 
and to all the other values of $q^2$ by means of a 
rigorous analytic continuation technique, based on dispersion relations.
The free parameters of such a description, which cover in principle
the whole $q^2$-complex plane, are determined by imposing theoretical 
and experimental constraints.
\subsection{Extracting the transition form factor from data} 
\label{tff-data}
To extract experimental values of the tff's, data on decay rates and cross 
section are compared with the kinematic factors computed in the
case of constant couplings, i.e. pointlike mesons [eqs.~(\ref{k-g}-\ref{k-s})].
The explicit forms for the conversion currents needed to compute the
kinematic factors in the cases: $M=$ pseudoscalar ($P$) and
$M=$ scalar ($S$), are [eq.~(\ref{j-ab})]:
\be
J_{\phi M}^\mu=\mathcal{J}_{\phi M}^\mu \cdot F_{\phi M}(s)
\hspace{20mm}\begin{array}{l}
\mathcal{J}_{\phi P}^\mu=e\,\epsilon^{\mu\nu\rho\sigma}\ve_\nu p_\rho q_\sigma\\
\\
\mathcal{J}_{\phi S}^\mu=e\,\left[p^\mu q^\nu-g^{\mu\nu}(pq)\right]\ve_\nu\\
\end{array},
\label{tensors}
\en
where $\epsilon^{\mu\nu\rho\sigma}$ is the fully antisymmetric Levi-Civita
tensor and, following the labeling of fig.~\ref{j-ab-fig}, $q$ is the 4-momentum 
of the photon and, $p$ and $\ve$ are the 4-momentum and the polarization 
vector of the $\phi$.
\bfi[ht!]
\bc
\epsfig{file=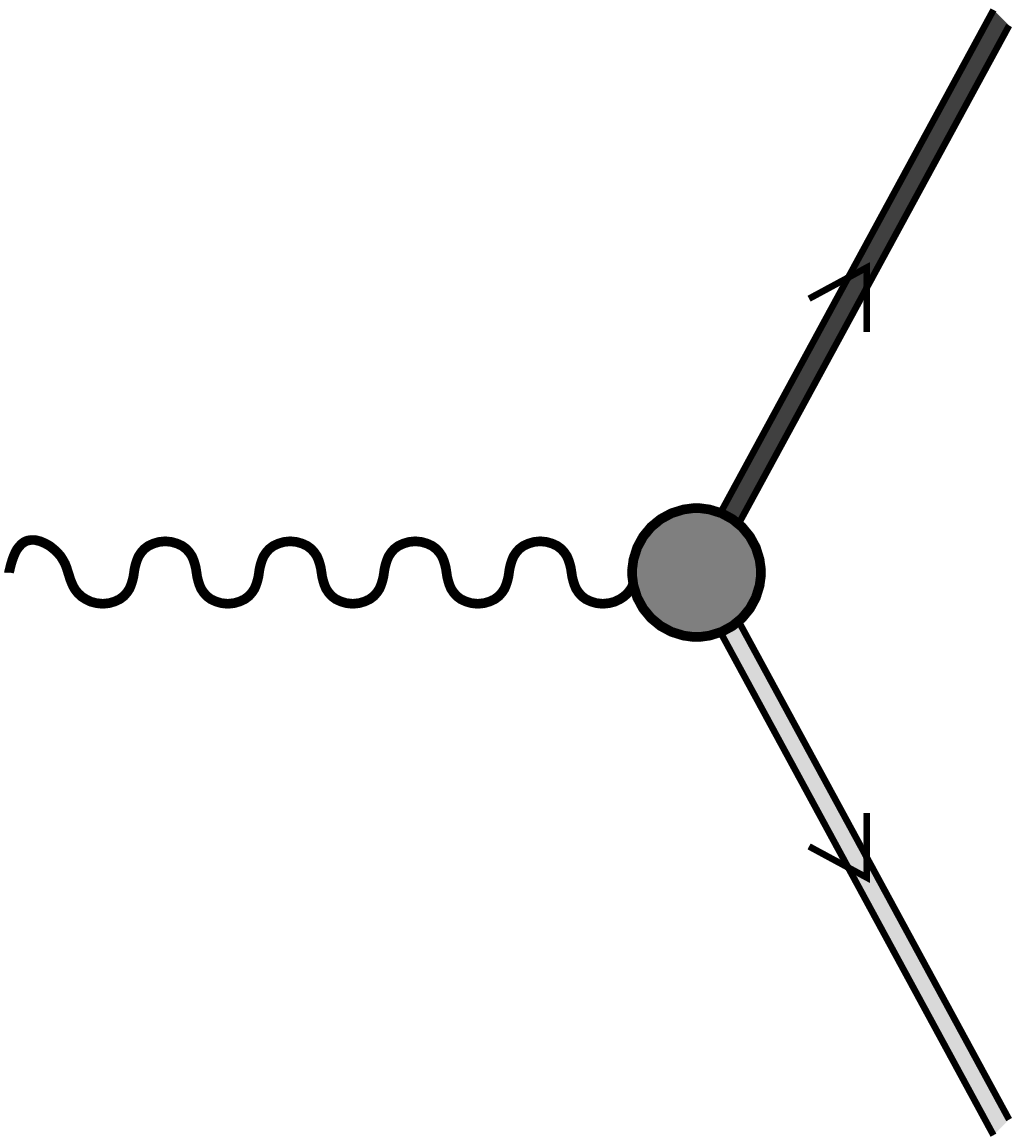,height=30mm}
\put(-98,42){$\gamma(q)$}
\put(-0,-2){$P,S(k)$}
\put(-0,80){$\phi(p,\ve)$}
\put(-15,38){$F_{\phi M}(q^2)$}
\caption{\label{j-ab-fig}
Diagram for the conversion $\phi\gamma P,S$.
}
\ec
\efi\\
The structure of the tensors which describe the dynamics of the
pseudoscalar (scalar) meson in the vertex $\phi\gamma P(S)$, reported
in eq.~(\ref{tensors}), follows from general arguments, i.e. spin properties,
Lorentz and gauge invariance. In particular, we note that in both cases,
only one tff is needed to describe the conversion.
\\
Once the conversion currents have been defined, rates and cross section may
be computed. First, we consider the radiative decay rates: $\phi(p)\ra M(k)\gamma(q)$
shown in fig.~\ref{agb}$a$ with $A\equiv\phi$ and $B\equiv M=P,S$, the
Feynman amplitudes read:
\be
\mathcal{M}(\phi\ra P\gamma)\!&=&\!eF_{\phi P}(0)\epsilon^{\mu\nu\rho\sigma} 
\ve_\nu p_\rho q_\sigma  {\ve'}^*_\mu\no\\
&&\label{amp-rad}\\
\mathcal{M}(\phi\ra S\gamma)\!&=&\!eF_{\phi S}(0)[p^\mu q^\nu-g^{\mu\nu}(pq)]
\ve_\nu{\ve'}^*_\mu\no,
\en
where $\ve'$ is the photon polarization vector. Both these amplitudes depend on
the value of the tff at $q^2=0$ because the photon is real. 
From these, we calculate the radiative decay rates:
\be
\Gamma(\phi\ra P\gamma)\!&=&\!\frac{\alpha}{3}\left(\frac{M_\phi^2-M_P^2}{2M_\phi}\right)^3
[F_{\phi P}(0)]^2\no\\
&&\label{rate-rad}\\
\Gamma(\phi\ra S\gamma)\!&=&\!\frac{\alpha}{3}\left(\frac{M_\phi^2-M_S^2}{2M_\phi}\right)^3
[F_{\phi S}(0)]^2\no.
\en
The kinematic factors are exactly the same, hence, at $q^2=0$, the different
natures of the mesons are not kinematically distinguishable.\\
The amplitudes for the conversion decays $\phi(p)\ra P,S(k)e^+(p_+)e^-(p_-)$
in the Born approximation, shown in fig.~\ref{agb}$b$ ($A\equiv\phi$ and 
$B\equiv M=P,S$), are:
\be
\mathcal{M}(\phi\ra Pe^+e^-)\!&=&\!-ie^2\big[F_{\phi P}(q^2)\epsilon^{\mu\nu\rho\sigma}
\ve_\nu p_\rho q_\sigma \big]\,\frac{1}{q^2}\,\big[\ov{u}(p_-)\gamma_\mu v(p_+)\big]\no\\
&&\label{amp-diff}\\
\mathcal{M}(\phi\ra Se^+e^-)\!&=&\!-ie^2\big\{F_{\phi S}(q^2)[p^\mu q^\nu-g^{\mu\nu}(pq)]\ve_\nu\big\}
\,\frac{1}{q^2}\,\big[\ov{u}(p_-)\gamma_\mu v(p_+)\big]\no,
\en
with: $(2m_e)^2<q^2<(M_\phi-M_M)^2$, and the corresponding
differential decay rates:
\be
\frac{d\Gamma(\phi\ra Pe^+e^-)}{dq^2}\!&=&\!
\frac{\alpha^2}{9\pi}\frac{1}{(2M_\phi)^3}\sqrt{1\!-\!\frac{4m_e^2}{q^2}}
\left(1\!+\!\frac{2m_e^2}{q^2}\right)
\frac{1}{q^2}\times\no\\
&&\!\!\times
\Big[\big(q^2\!+\!M_\phi^2\!-\!M_P^2\big)^2\!-\!4M_\phi^2 q^2\Big]^\frac{3}{2}
|F_{\phi P}(q^2)|^2
\no\\
&&\label{rate-diff}\\
\frac{d\Gamma(\phi\ra Se^+e^-)}{dq^2}\!&=&\!
\frac{\alpha^2}{9\pi}\frac{1}{(2M_\phi)^3}\sqrt{1\!-\!\frac{4m_e^2}{q^2}}
\left(1\!+\!\frac{2m_e^2}{q^2}\right)
\frac{1}{q^2}\times\no\\
&&\!\!\times
\Big[\big(q^2\!+\!M_\phi^2\!-\!M_S^2\big)^2\!+\!2M_\phi^2 q^2\Big]
\Big[\big(q^2\!+\!M_\phi^2\!-\!M_S^2\big)^2\!-\!4M_\phi^2 q^2\Big]^\frac{1}{2}
|F_{\phi S}(q^2)|^2\no.
\en
In this $q^2$ region the different natures of the meson $M$ provide
different kinematic structures. \\
Finally we consider the annihilation process $e^+(p_+)e^-(p_-)\ra \phi(p) P,S(k)$, 
shown in fig.~\ref{agb}$c$ ($A\equiv\phi$ and $B\equiv M=P,S$) in the Born
approximation. The amplitudes are:
\be
\mathcal{M}(e^+e^-\ra\phi P)\!&=&\!-ie^2\big[F_{\phi P}(q^2)\epsilon^{\mu\nu\rho\sigma}
\ve_\nu p_\rho q_\sigma \big]\,\frac{1}{q^2}\,\big[\ov{v}(p_+)\gamma_\mu u(p_-)\big]\no\\
&&\label{amp-annihi}\\
\mathcal{M}(e^+e^-\ra\phi S)\!&=&\!-ie^2\big\{F_{\phi S}(q^2)[p^\mu q^\nu-g^{\mu\nu}(pq)]\ve_\nu\big\}
\,\frac{1}{q^2}\,\big[\ov{v}(p_+)\gamma_\mu u(p_-)\big]\no,
\en
for $q^2>(M_\phi+M_M)^2$ and the cross sections:
\be
\sigma(e^+e^-\ra\phi P)\!&=&\!\frac{\pi}{6}\frac{\alpha^2}{(q^2)^3}
\frac{q^2+2m_e^2}{\sqrt{q^2(q^2-4m_e^2)}}\times\no\\
&&\!\!\times
\Big[\big(q^2\!+\!M_\phi^2\!-\!M_P^2\big)^2\!-\!4M_\phi^2 q^2\Big]^\frac{3}{2}
|F_{\phi P}(q^2)|^2\no\\
&&\label{rate-annihi}\\
\sigma(e^+e^-\ra\phi S)\!&=&\!\frac{\pi}{6}\frac{\alpha^2}{(q^2)^3}
\frac{q^2+2m_e^2}{\sqrt{q^2(q^2-4m_e^2)}}\times\no\\
&&\!\!\times
\Big[\big(q^2\!+\!M_\phi^2\!-\!M_S^2\big)^2\!+\!2M_\phi^2 q^2\Big]
\Big[\big(q^2\!+\!M_\phi^2\!-\!M_S^2\big)^2\!-\!4M_\phi^2 q^2\Big]^\frac{1}{2}
|F_{\phi S}(q^2)|^2\no.
\en
Once again the pseudoscalar and scalar natures 
of the meson $M$ are reflected in the different kinematic factors that have 
to be considered to extract tff's from data.\\
It is interesting to note that the dimension of the tensors~(\ref{tensors}) 
is not that of a current, i.e. energy. It follows that the tff's, unlike
static ff's, are dimensional quantities. In particular, we have:
\be
\left[\mathcal{J}_{\phi M}^\mu\right]=E^2\hspace{10mm}
\Longrightarrow
\hspace{10mm} \left[F_{\phi M}\right]=E^{-1}.
\label{dimensions}
\en
This is why, when we are dealing with tff's, the kinematic
factor which multiplies the tff itself to give the physical quantities: decay rates or cross sections, 
cannot be interpreted as the pointlike physical quantities. To get pointlike quantities we must
replace the tff's with dimensional ($E^{-1}$) coupling constants.
Such a replacement should provide an unusual result in case of 
annihilation cross section. If we consider
the formulae of eq.~(\ref{rate-annihi}), where we put
$F_{\phi M}(q^2)\equiv g^\gamma_{\phi M}$ ($g^\gamma_{\phi M}$ is a
constant coupling with: $\left[g^\gamma_{\phi M}\right]=E^{-1}$), in the limit 
$q^2\ra \infty$, we get:
\be
\sigma(e^+e^-\ra\phi P)\!&&\!\mathop{\longrightarrow}_{q^2\ra\infty} 
\frac{\pi\alpha^2}{6}|g^\gamma_{\phi P}|^2
\no\\
\sigma(e^+e^-\ra\phi S)\!&&\!\mathop{\longrightarrow}_{q^2\ra\infty} 
\frac{\pi\alpha^2}{6}|g^\gamma_{\phi S}|^2.
\en
Constant asymptotic limits, in case of pointlike mesons, represent
an unexpected result. The usual ``minimal'' expectation
is a behavior that mimics the cross section
$\sigma(e^+e^-\ra \mu^+\mu^-)$ which vanishes like $(1/q^2)$ as $q^2$ 
diverges. However, such expected behavior is true
only in case of the static ff where we have nonvanishing amplitudes only in
the case of charged mesons. Since we are dealing with pairs of different neutral 
mesons which convert from one in to the other and since, even for ``pointlike'' 
mesons, we must consider the quark structure of these particles which allows
the conversion (dynamic ff's), then additional effects
providing further attenuation, must be accounted for. 
As we will see in the following, attenuation factors
in the asymptotic regime are provided by the hadronic helicity 
rule~\cite{brodsky}.
\subsection{Parameterization in the resonance region}
\label{res-reg}
The resonance region covers the portion of the time-like region where
the tff is characterized by the vector meson resonance contributions.
In general this interval goes from the theoretical threshold $s_{\rm th}$
up to $s_{\rm asy}\sim (3-4\;{\rm GeV})^2$. Above this energy, the power
law asymptotic behavior, as predicted by QcR~\cite{brodsky}, is assumed. 
Vector meson resonances, in the Born approximation, may
be interpreted as intermediate states coupling the virtual photon and final mesons.
They carry the quantum numbers of the photon, i.e.
$J^{PC}=1^{--}$. In light of this, and following the definition~(\ref{tensors}), 
the conversion current $\phi\gamma M_I$, where $I$ is the isospin of the 
meson $M$, has the form:
\be
J_{\phi M_I}^\mu=\mathcal{J}_{\phi M_I}^\mu \sum_{j=1}^{N_{M_I}}
\frac{M_{j}^2}{eF_{V_j}}\,
\frac{g^{V_j}_{\phi M_I}}{M_{j}^2-q^2-i\Gamma_j M_j},
\label{eq:corr}
\en
where, as shown in fig.~\ref{sum-v}, $N_{M_I}$ intermediate vector mesons $V_j$, 
with mass $M_j$ and width $\Gamma_j$, are considered. 
\bfi[h!]
\bc\vspace{0mm}
\epsfig{file=int-v.eps,height=30mm}
\put(-83,42){$\gamma$}
\put(-0,-2){$M_I$}
\put(-0,80){$\phi$}
\put(-15,38){$F_{\phi M_I}$}
\hspace{20mm}
\epsfig{file=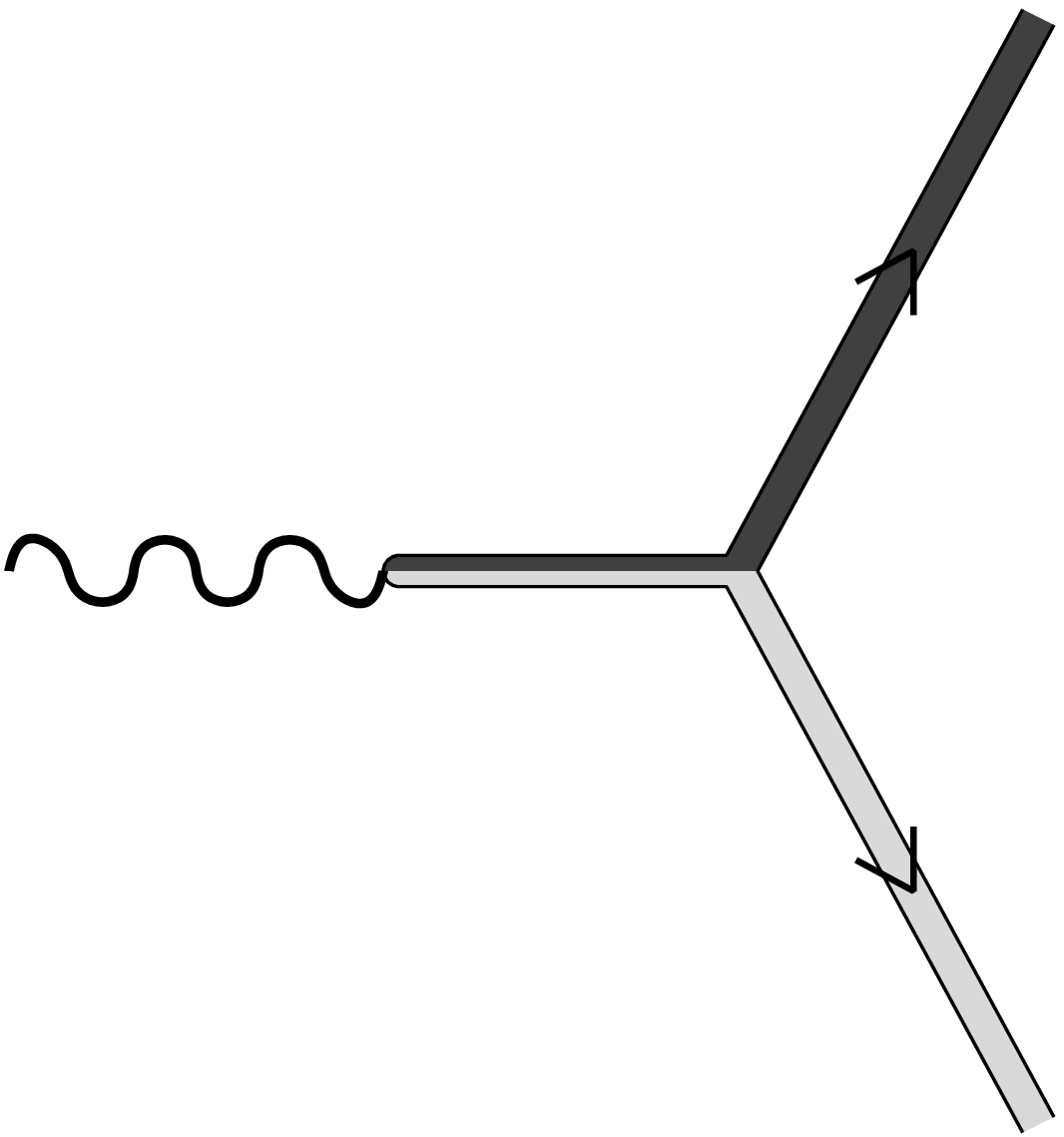,height=30mm}
\put(-120,42){$=\,\sum\limits_{j=1}^{N_{M_I}}$}
\put(-42,48){$V_j$}
\put(-68,26){
\bm{10mm}
$$
\frac{M_{V_j}^2}{F_{V_j}}
$$
\em
}
\put(-87,42){$\gamma$}
\put(-0,-2){$M_I$}
\put(-0,80){$\phi$}
\put(-15,38){$g^{V_j}_{\phi M_I}$}
\vspace{-5mm}
\ec
\caption{\label{sum-v}
Schematic representation of the parameterization for tff's.
}
\efi\\
To summarize, by assuming that the virtual photon couples with all the allowed
vector mesons $V_j$ with a strength $\frac{M_{V_j}^2}{F_{V_j}}$ and then
each meson $V_j$ goes in the final state $\phi M_I$ with a coupling 
$g^{V_j}_{\phi M_I}$, we get the tff:
\be
F_{\phi M_I}(q^2)=\sum_{j=1}^{N_{M_I}} \frac{M_j^2}{eF_{V_j}}\,
\frac{g^{V_j}_{\phi M_I}}{M_j^2-q^2-i\Gamma_j M_j}
\hspace{20mm} s_{\rm th}^{I}\le q^2\le s_{\rm asy}^{M_I}.
\label{tff-res}
\en
Note that the species which contribute to this sum and also the
threshold values $s_{\rm th}^{I}$ and $s_{\rm asy}^{M_I}$ depend 
on the nature of the meson $M_I$.
%
In particular, while $s_{\rm asy}^{M_I}$ is a free parameter of the fit, 
$s_{\rm th}^{I}$ depends on the isospin $I$ of the meson $M_I$, i.e.:
\be
s_{\rm th}^{I}=\left\{
\begin{array}{ll}
(3M_\pi)^2 & I=0\\
(2M_\pi)^2 & I=1\\
\end{array}
\right. .
\label{sth}
\en
This parameterization integrates all the information about
the structure of the meson $M_I$\footnotemark[3].%
\footnotetext[3]{%
E.g., for $M_0\equiv\f$, a strong affinity of the $\f$ with
the $K\ov{K}$ intermediate state (kaon loop) should manifest itself
in an enhancement of the coupling $g^\phi_{\phi\f}$, since in the $s$-channel
the $K\ov{K}$ state is almost completely resonant in $\phi(1020)$.}
%
%
\subsection{Selection of the resonant contributions}
\label{select}
\subsubsection{Isoscalar $M_{I=0}$}
\label{select-im0}
The first criterion used to select vector meson contributions to the tff's
in the resonance region, where we adopt the parameterization~(\ref{tff-res}), 
is the quantum number conservation.
If we consider a pseudoscalar [$I^G(J^{PC})=0^+(1^{-+})$] and a scalar 
[$I^G(J^{PC})=0^+(0^{++})$] meson with isospin zero, having the 
$\phi$ quantum numbers: $I^G(J^{PC})=0^-(1^{--})$, both $\phi P_0$ and $\phi S_0$
final states (the subscript $0$ indicates the isospin) 
will have $I^G(J^{PC})=0^-(1^{--})$, hence only contributions from the
$\omega$- and $\phi$-family are expected. However, by considering explicitly 
the structure of the mesons in terms of valence quarks, using only the light
quarks $q=u,d,s$ in a $q\ov{q}$ bound state, we have:
\be
|\phi\rangle= s\ov{s}
\hspace{30mm}
\begin{array}{l}
|P_0\rangle= X_{P_0}|u,d;-\rangle+Y_{P_0}|s,s;-\rangle\\
|S_0\rangle= X_{S_0}|u,d;+\rangle+Y_{S_0}|s,s;+\rangle\\
\end{array}
\label{q-comp}
\en 
with the normalization: $X_{P_0,S_0}^2+Y_{P_0,S_0}^2=1$ and 
where the state $|q_1,q_2;P\rangle$ is a normalized
combination of $q_1\ov{q}_1$ and $q_2\ov{q}_2$, with
parity $P$ and $J^{C}=0^+$.
While the $\phi$-family contributions are allowed (fig.~\ref{phi-eta}),
the $\omega$-family contributions are instead OZI-forbidden\cite{ozi}.
In fact, as it is shown in fig.~\ref{om-eta}, the corresponding Feynman 
diagrams have disconnected flavor lines in the vector sector.\\
It follows that, in light of the quantum number conservation and the
OZI rule, only $\phi$-family contributions are expected for
both $\phi P_0$ and $\phi S_0$ final states.
\bfi[ht]
%
\hspace{15mm}
\epsfig{file=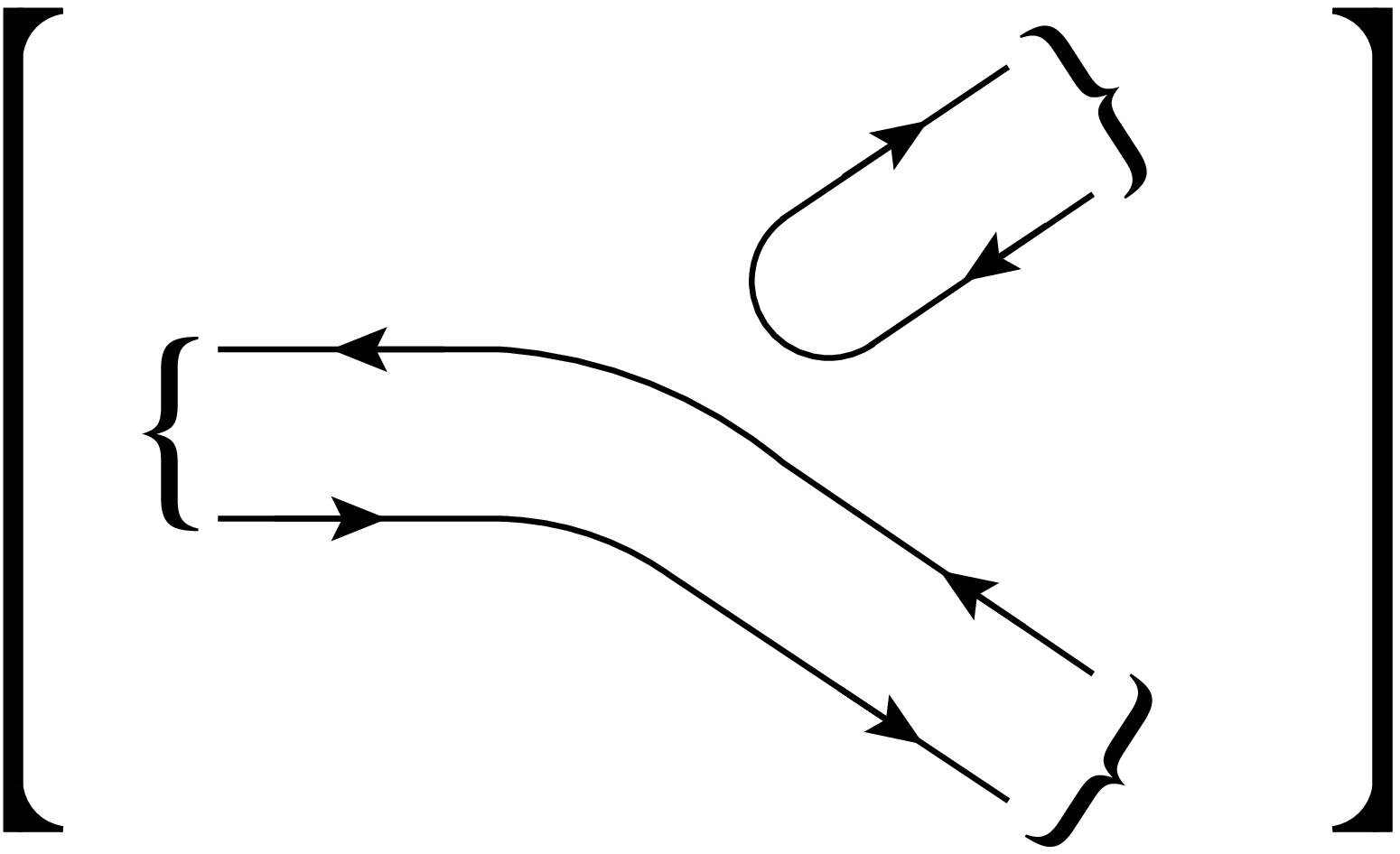,width=56.17mm}
\put(-133,65.5){$\ov{s}$}
\put(-133,45.5){$s$}
\put(-55,74){\rotatebox{36}{$\ov{u},\ov{d}$}}
\put(-64,89){\rotatebox{36}{$u,d$}}
\put(-38,30){\rotatebox{-36}{$\ov{s}$}}
\put(-47,16){\rotatebox{-36}{$s$}}
\put(-29,10){$\phi$}
\put(-32.5,92){$|u\!,\!d;\!\!\mp\!\rangle$}
\put(-155,50){\rotatebox{90}{$\phi^*$}}
\put(-195,50){\large$X_{P_0(S_0)}$}
\hspace{15mm}
\epsfig{file=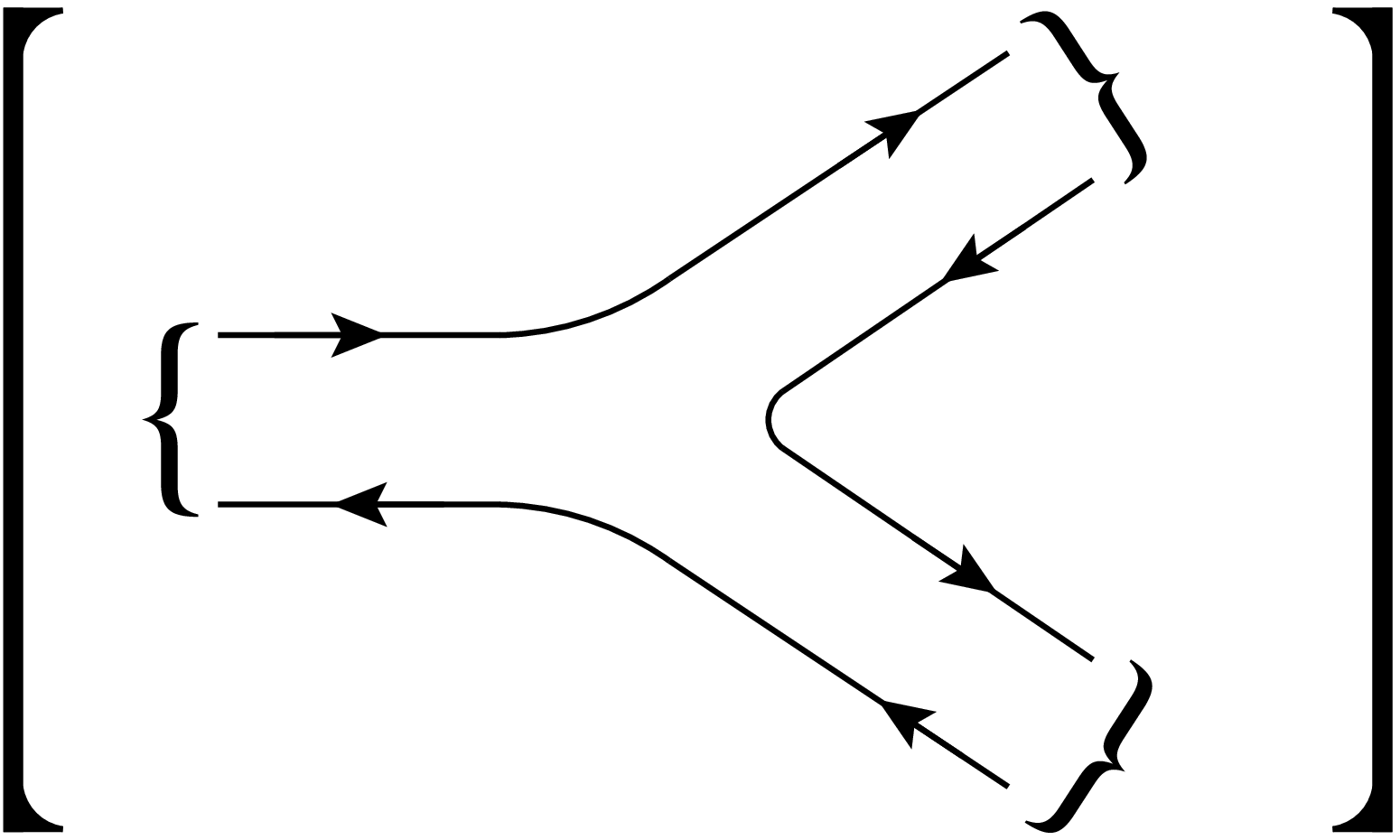,width=56.17mm}
\put(-133,65.5){$s$}
\put(-133,45.5){$\ov{s}$}
\put(-44,80){\rotatebox{36}{$\ov{s}$}}
\put(-53,94){\rotatebox{36}{$s$}}
\put(-38,30){\rotatebox{-36}{$s$}}
\put(-47,16){\rotatebox{-36}{$\ov{s}$}}
\put(-29,10){$\phi$}
\put(-32.5,92){$|s\!,\!s;\!\!\mp\!\rangle$}
\put(-155,50){\rotatebox{90}{$\phi^*$}}
\put(-193,50){\large$Y_{P_0(S_0)}$}
\put(-203,50){\large$+$}
\caption{\label{phi-eta}
Feynman diagram for $\phi^*\ra\phi P_0(S_0)$ with: $|\phi\rangle= s\ov{s}$ and
$|P_0(S_0)\rangle= X_{P_0,S_0}|u,d;\mp\rangle+Y_{P_0,S_0}|s,s;\mp\rangle$.
The two components $|u,d;\mp\rangle$ and $|s,s;\mp\rangle$ are separately shown.
}
\efi
\bfi[ht]\vspace{-0mm}
\hspace{15mm}
\epsfig{file=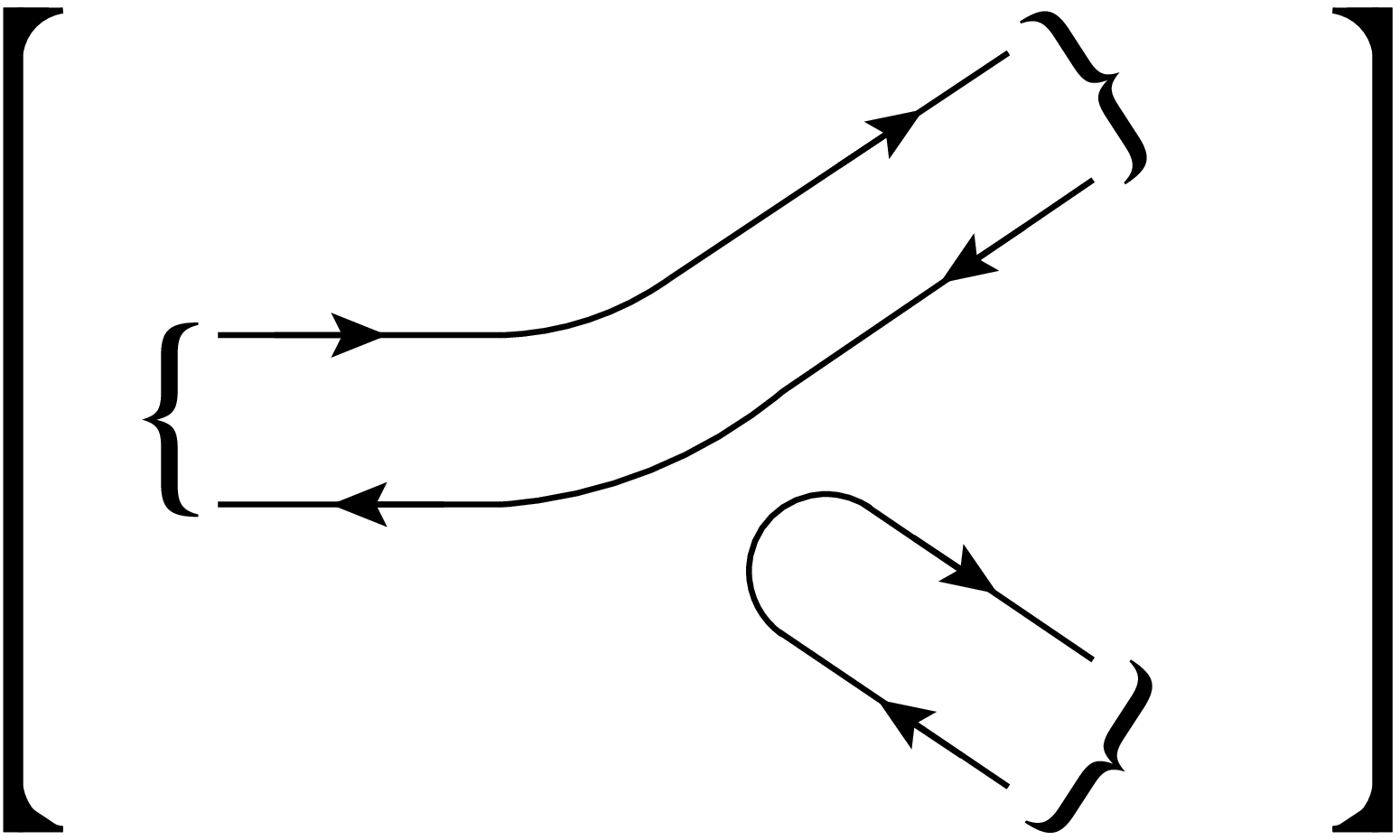,width=56.17mm}
\put(-133,66){$u,d$}
\put(-133,46){$\ov{u},\ov{d}$}
\put(-55,74){\rotatebox{36}{$\ov{u},\ov{d}$}}
\put(-64,89){\rotatebox{36}{$u,d$}}
\put(-38,30){\rotatebox{-36}{$s$}}
\put(-47,16){\rotatebox{-36}{$\ov{s}$}}
\put(-29,10){$\phi$}
\put(-32.5,92){$|u\!,\!d;\!\!\mp\!\rangle$}
\put(-154,50){\rotatebox{90}{$\omega^*$}}
\put(-195,50){\large$X_{P_0(S_0)}$}
\hspace{15mm}
\epsfig{file=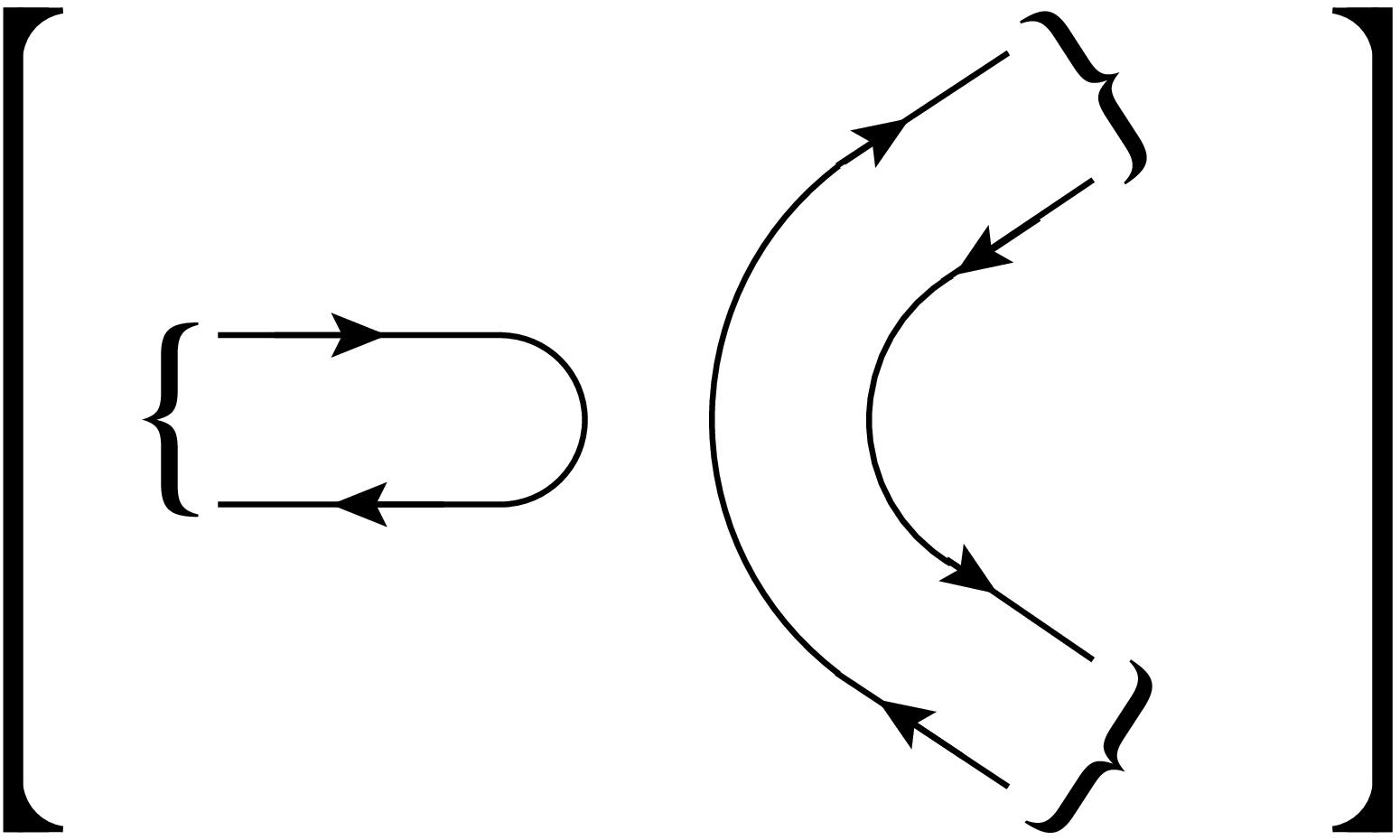,width=56.17mm}
\put(-133,66){$u,d$}
\put(-133,46){$\ov{u},\ov{d}$}
\put(-44,80){\rotatebox{36}{$\ov{s}$}}
\put(-53,94){\rotatebox{36}{$s$}}
\put(-38,30){\rotatebox{-36}{$s$}}
\put(-47,16){\rotatebox{-36}{$\ov{s}$}}
\put(-29,10){$\phi$}
\put(-31.5,92){$|s\!,\!s;\!\!\mp\!\rangle$}
\put(-154,50){\rotatebox{90}{$\omega^*$}}
\put(-193,50){\large$Y_{P_0(S_0)}$}
\put(-203,50){\large$+$}
\caption{\label{om-eta}
Feynman diagram for $\omega^*\ra\phi P_0(S_0)$ with: $|\phi\rangle= s\ov{s}$ and
$|P_0(S_0)\rangle= X_{P_0,S_0}|u,d;\mp\rangle+Y_{P_0,S_0}|s,s;\mp\rangle$.
The two components $|u,d;\mp\rangle$ and $|s,s;\mp\rangle$ are separately shown.
}
\efi\\
%
%
%
Another interesting hypothesis is that the meson $M_0$ is a scalar tetraquark 
$S_0^4$\cite{4q}. In the most stable configuration, the four quarks should 
arrange in diquark-antidiquark pairs $[qq][\ov{qq}]$ (in each diquark the two quarks are bound 
in $S$-wave to form a color and flavor $\ov{3}$ with $J^P=0^+$, totally antisymmetric
in color, flavor and spin), with zero total spin and angular momentum. 
The quantum numbers of the $\phi S_0^4$ final state remain $I^G(J^{PC})=0^-(1^{--})$, 
but, if we assume a structure like $[ns][\ov{ns}]$, where $n=u,d$, for the meson $S_0^4$, 
then both $\phi$- and $\omega$-family contributions are now OZI-favored (see fig.~\ref{new-hyp-fig}).
\bfi[ht]
\bc
\epsfig{file=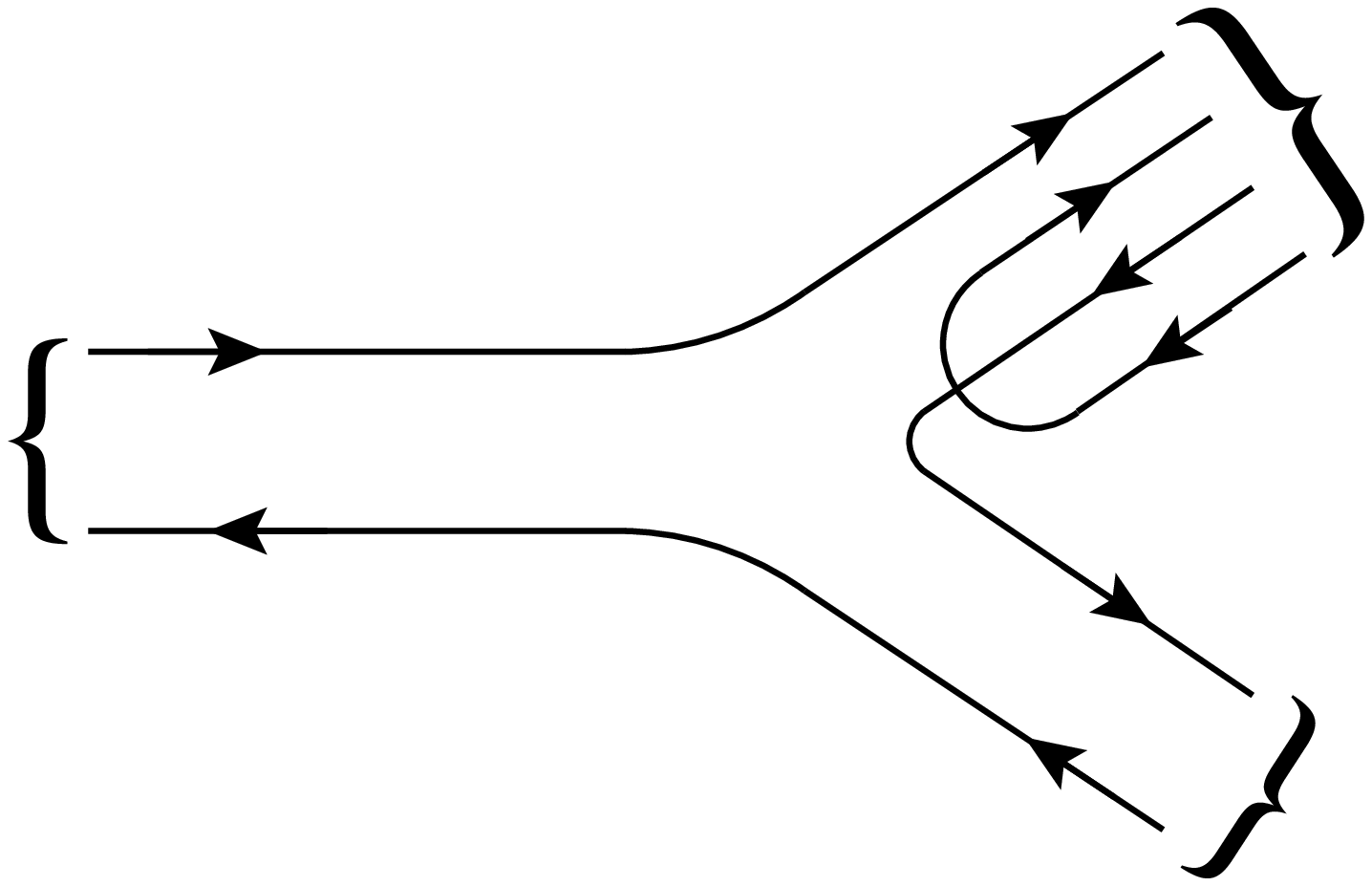,width=65mm}
\put(-170,75){$s$}
\put(-170,52){$\ov{s}$}
\put(-194,59){$\phi^*$}
\put(-46,96.5){\scriptsize\rotatebox{33}{$u,d$}}
\put(-35,79){\scriptsize\rotatebox{33}{$\ov{u},\ov{d}$}}
\put(-32,93){\scriptsize\rotatebox{33}{$\ov{s}$}}
\put(-42.,110.5){\scriptsize\rotatebox{33}{$s$}}
\put(-13,104){$S_0^4$}
\put(-38,16.5){\scriptsize\rotatebox{-33}{$\ov{s}$}}
\put(-27.,33.5){\scriptsize\rotatebox{-33}{$s$}}
\put(-18,11){$\phi$}
\hspace{8mm}
\epsfig{file=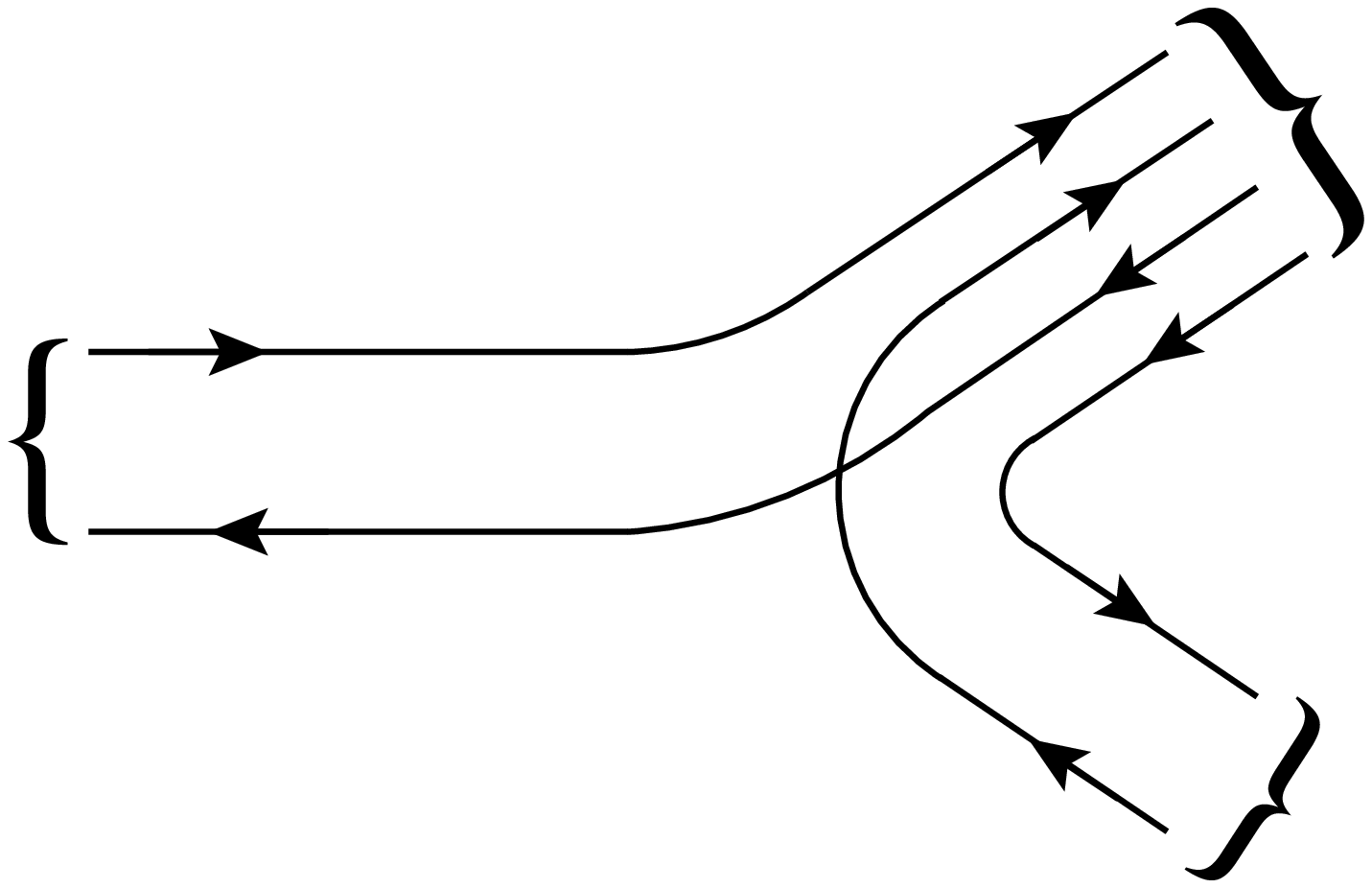,width=65mm}
\put(-170,76){$u,d$}
\put(-170,53){$\ov{u},\ov{d}$}
\put(-194,59){$\omega^*$}
\put(-51,105.5){\scriptsize\rotatebox{33}{$u,d$}}
\put(-41,88){\scriptsize\rotatebox{33}{$\ov{u},\ov{d}$}}
\put(-25.,85){\scriptsize\rotatebox{33}{$\ov{s}$}}
\put(-36.,101.5){\scriptsize\rotatebox{33}{$s$}}
\put(-13,104){$S_0^4$}
\put(-38,16.5){\scriptsize\rotatebox{-33}{$\ov{s}$}}
\put(-27.,33.5){\scriptsize\rotatebox{-33}{$s$}}
\put(-18,11){$\phi$}
\caption{\label{new-hyp-fig}
Feynman diagrams for the $\phi^*$ and $\omega^*$ contribution
to the $\phi S_0^4$ final state.
}
\ec
\efi\vspace{-5mm}\\
To summarize, in the resonance region we consider three possibilities corresponding to 
different choices of the meson $M_0$:
\bi
\item $M_0=$ isospin zero, pseudoscalar meson $P_0$:
      \be
      F_{\phi P_0}^{\rm Res}(q)=\sum\limits_{V=\phi,\phi',\ldots} \frac{M_V^2}{eF_{V}}
      g^{V}_{\phi P_0}\frac{e^{i\Phi_{VP_0}}}{M_V^2-q^2-i\Gamma_V M_V}
      \hspace{15mm} s_{\rm th}^0\le q^2\le s_{\rm asy}^{P_0};
      \label{res-p}
      \en
\item $M_0=$ isospin zero, scalar $q\ov{q}$ meson $S_0$:
      \be
      F_{\phi S_0}^{\rm Res}(q)=\sum\limits_{V=\phi,\phi',\ldots} \frac{M_V^2}{eF_{V}}
      g^{V}_{\phi S_0}\frac{e^{i\Phi_{VS_0}}}{M_V^2-q^2-i\Gamma_V M_V}
      \hspace{15mm} s_{\rm th}^0\le q^2\le s_{\rm asy}^{S_0};
      \label{res-s}
      \en
\item $M_0=$ isospin zero, scalar tetraquark meson $S_0^4$:
      \be
      F_{\phi S_0^4}^{\rm Res}(q)=\sum\limits_{V=\phi,\phi',\ldots,\omega,\omega',\ldots} 
      \frac{M_V^2}{eF_{V}}
      g^{V}_{\phi S_0^4}\frac{e^{i\Phi_{VS_0^4}}}{M_V^2-q^2-i\Gamma_V M_V}
      \hspace{5mm} s_{\rm th}^0\le q^2\le s_{\rm asy}^{S_0^4}.
      \label{res-s4}
      \en
\ei
Relative phases $\Phi_{V M_0}$ have been included to account for possible re-scattering
effects.
In all these cases the isospin of the final state is zero, hence the theoretical threshold is
$s_{\rm th}^0=(3M_\pi)^2$ [eq.~(\ref{sth})].
\subsubsection{Isovector $M_{I=1}$}
\label{select-im1}
In case of isovector pseudoscalar [$I^G(J^{PC})=1^-(0^{-+})$] 
or scalar [$I^G(J^{PC})=1^-(0^{++})$] meson $M_1$ the final state $\phi M_1$
(the subscript $1$ is the isospin) has quantum numbers $I^G(J^{PC})=1^+(1^{--})$.
It follows that only $\rho$-family contributions are expected. However,
from the Feynman diagrams shown in fig.~\ref{rho-pi0}, constructed 
using for $\phi$, $P_1$, and $S_1$ the quark structures of eq.~(\ref{q-comp}),
we note that all these contributions are OZI-suppressed.
\bfi[ht]\vspace{-0mm}
\hspace{15mm}
\epsfig{file=om-eta-phi1-new.eps,width=56.17mm}
\put(-133,66){$u,d$}
\put(-133,46){$\ov{u},\ov{d}$}
\put(-55,74){\rotatebox{36}{$\ov{u},\ov{d}$}}
\put(-64,89){\rotatebox{36}{$u,d$}}
\put(-38,30){\rotatebox{-36}{$s$}}
\put(-47,16){\rotatebox{-36}{$\ov{s}$}}
\put(-29,10){$\phi$}
\put(-32.5,92){$|u\!,\!d;\!\!\mp\!\rangle$}
\put(-154,50){\rotatebox{90}{$\rho^*$}}
\put(-195,50){\large$X_{P_1(S_1)}$}
\hspace{13mm}
\epsfig{file=om-eta-phi2-new.eps,width=56.17mm}
\put(-133,66){$u,d$}
\put(-133,46){$\ov{u},\ov{d}$}
\put(-44,80){\rotatebox{36}{$\ov{s}$}}
\put(-53,94){\rotatebox{36}{$s$}}
\put(-38,30){\rotatebox{-36}{$s$}}
\put(-47,16){\rotatebox{-36}{$\ov{s}$}}
\put(-29,10){$\phi$}
\put(-32.5,92){$|s\!,\!s;\!\!\mp\!\rangle$}
\put(-154,50){\rotatebox{90}{$\rho^*$}}
\put(-193,50){\large$Y_{P_1(S_1)}$}
\put(-201,50){\large$+$}
\caption{\label{rho-pi0}
Feynman diagram for $\rho^*\ra\phi P_1(S_1)$ with: $|\phi\rangle= s\ov{s}$ and
$|P_1(S_1)\rangle= X_{P_1,S_1}|u,d;\mp\rangle+Y_{P_1,S_1}|s,s;\mp\rangle$.
The two components $|u,d;\mp\rangle$ and $|s,s;\mp\rangle$ are separately shown.
}
\efi\\
If we consider an isovector tetraquark scalar meson $S_1^4$, with structure
$[ns][\ov{ns}]$ ($n=u,d$ as in case of isospin zero), then the additional
hadronic field leads to OZI-favored coupling between intermediate $\rho$
recurrences and the final state $\phi S_1^4$ (see fig.~\ref{s4-i1}).
\bfi[ht]
\bc
\epsfig{file=om-f0-phi1-new.eps,width=65mm}
\put(-170,76){$u,d$}
\put(-170,53){$\ov{u},\ov{d}$}
\put(-194,59){$\rho^*$}
\put(-51,105.5){\scriptsize\rotatebox{33}{$u,d$}}
\put(-41,88){\scriptsize\rotatebox{33}{$\ov{u},\ov{d}$}}
\put(-25.,85){\scriptsize\rotatebox{33}{$\ov{s}$}}
\put(-36.,101.5){\scriptsize\rotatebox{33}{$s$}}
\put(-13,104){$S_1^4$}
\put(-38,16.5){\scriptsize\rotatebox{-33}{$\ov{s}$}}
\put(-27.,33.5){\scriptsize\rotatebox{-33}{$s$}}
\put(-18,11){$\phi$}
\caption{\label{s4-i1}
Feynman diagrams for the $\rho^*$ contribution
to the $\phi S_1^4$ final state.
}
\ec
\efi
\\
To summarize, in the resonance region, two possible choices for $M_1$ are considered:
\bi
\item $M_1=$ isospin one, pseudoscalar  $P_1$ or scalar meson $S_1$:
      \be
      F_{\phi M_1}^{\rm Res}(q)=\sum\limits_{V=\rho,\rho',\ldots} \frac{M_V^2}{eF_{V}}
      g^{V}_{\phi M_1}\frac{e^{i\Phi_{VM_1}}}{M_V^2-q^2-i\Gamma_V M_V}
      \hspace{15mm} s_{\rm th}^1\le q^2\le s_{\rm asy}^{M_1},
      \label{res-ps1}
      \en
      with $M_1=P_1,S_1$ and for all the couplings holds $g^{V}_{\phi M_1}\sim 0$,
      due to the OZI suppression (see fig.~\ref{rho-pi0});
\item $M_1=$ isospin one, scalar tetraquark meson $S_1^4$:
      \be
      F_{\phi S_1^4}^{\rm Res}(q)=\sum\limits_{V=\rho,\rho',\ldots}
      \frac{M_V^2}{eF_{V}}
      g^{V}_{\phi S_1^4}\frac{e^{i\Phi_{VS_1^4}}}{M_V^2-q^2-i\Gamma_V M_V}
      \hspace{10mm} s_{\rm th}^1\le q^2\le s_{\rm asy}^{S_1^4}.
      \label{res-s41}
      \en
\ei
Again, the relative phases $\Phi_{V M_1}$ account for possible re-scattering effects and, 
having isospin one, the theoretical threshold is now
$s_{\rm th}^1=(2M_\pi)^2$.
%
%
%
%
\subsection{The asymptotic behavior}
\label{asy-beha}
A key point of this procedure to determine the parameters of the vector resonances 
is the knowledge of the tff time-like asymptotic behavior. 
For this purpose, we adopt the nominal power law predicted
by QcR~\cite{brodsky}. It follows that, as $q^2$ diverges, 
a hadronic ff must vanish like $(1/q^2)$ to some power which linearly depends 
on the number of quark constituents and other quantum numbers of the hadrons 
under consideration. To describe the asymptotic behavior of the 
$\phi M_I$ tff's, we must also account for suppression factors related to
the quark structure of the mesons, i.e.: the violation of the hadronic 
helicity conservation\footnotemark[4]~\cite{hely}, which passes from zero, in the
leptonic initial state, to one in the $\phi M_I$ final state ($\lambda_{M_I}=0$, 
$|\lambda_\phi|=1$, where $\lambda_\alpha$ stands for the helicity of the particle $\alpha$), 
and the relative angular momentum between the quark and the antiquark in $M_I$. 
Since the asymptotic behavior does not depend on the isospin of the 
meson $M_I$, the following discussion holds for both $I=0$ and $I=1$.
%
%
\footnotetext[4]{%
The virtual photon of the annihilation $e^+e^-\ra\gamma^*\ra\phi M_I$, at high energies, 
always has spin $\pm 1$ along the beam axis. Angular momentum conservation implies that
the total angular momentum is 1. In the center of mass frame: 
$\vec{p}_\phi=-\vec{p}_{M_I}\equiv \vec{p}$, it follows that 
\be
\left|
\frac{\vec{p}_\phi\cdot\vec{s}_\phi}{|\vec{p}_\phi|}-
\frac{\vec{p}_{M_I}\cdot\vec{s}_{M_I}}{|\vec{p}_{M_I}|}
\right|=\left|\lambda_\phi-\lambda_{M_I}\right|=
\left|
\frac{\vec{p}\cdot\vec{s}_{\rm tot}}{|\vec{p}|}
\right|=0,1,
\label{hel}
\en
where $\vec{s}_\phi$ and $\vec{s}_{M_I}$, and $\lambda_\phi$ and $\lambda_{M_I}$ are 
the spins and the helicities of the final mesons. But, 
since the hadronic helicity conservation requires: $\lambda_\phi+\lambda_{M_I}=0$, or
$\lambda_\phi-\lambda_{M_I}=2\lambda_\phi=-2\lambda_{M_I}$, eq.~(\ref{hel}) holds, in case
of mesons, only if $|\lambda_\phi|=|\lambda_{M_I}|=0$. In the cases under consideration,
the coupling $\phi\gamma^* M_I$ is described in terms of only one tff 
$F_{\phi M_I}(q^2)$ and this requires $|\lambda_\phi|=1$. Helicity zero would mean
that the $\phi$ spin lies in the plane orthogonal to its 3-momentum, in this case
we would need an additional degree of freedom (for instance: the azimuthal angle) and then
an additional tff to describe this process. But since the tff is only one the $\phi$ spin 
must lie in 3-momentum direction, i.e. $|\lambda_\phi|=1$.
}
%
%
\\ 
As discussed in ref.~\cite{hely}, the cross section for the process $e^+e^-\to\phi M_I$
scaled by the $\mu^+\mu^-$ cross section goes asymptotically ($q^2\to\infty$) as:
\be
 \frac{\sigma_{\phi M_I}(q^2)}{\sigma_{\mu^+\mu^-}(q^2)}\propto 
\left(\frac{1}{q^2}\right)^{2(n_H-1)+n_\lambda+l_q}\hspace{7mm} 
\begin{array}{l}
n_H=\mbox{number of hadronic fields}\\
n_\lambda=|\lambda_{M_I}+\lambda_\phi|\\
l_q=q\ov{q}\mbox{ relative angular momentum in $M_I$.}\\
\end{array}
\en
The asymptotic behavior of $\sigma_{\phi M_I}(q^2)$ 
[eq.~(\ref{rate-annihi})] depends on the tff, while the one
of $\sigma_{\mu^+\mu^-}(q^2)$ is well-known:
\be
\sigma_{\phi M_I}(q^2)\mathop{\propto}_{q^2\ra\infty} |F_{\phi M_I}(q^2)|^2\hspace{20mm}
\sigma_{\mu^+\mu^-}(q^2)\mathop{\propto}_{q^2\ra\infty} \frac{1}{q^2}.
\en
From previous equations we may extract the asymptotic behavior
for the tff as:
\be
|F_{\phi M_I}(q^2)|\mathop{\propto}_{q^2\ra\infty}
\left(\frac{1}{q^2}\right)^{n_H+\frac{n_\lambda+l_q-1}{2}}.
\label{asy-gen}
\en
In the three cases investigated, we obtain the following asymptotic power laws:
\bi
\item for $M_I=P_I$: $(n_H,n_\lambda,l_q)=(2,1,0)$, there are two
      hadronic fields in the final state (fig.~\ref{phi-eta}), the hadronic
      helicity flips because $\lambda_{P_I}=0$ and $\lambda_\phi=1$ and,
      being the pseudoscalar $q\ov{q}$ state a $n^{2S+1}L_J=1^1S_0$
      element of the pseudoscalar flavor-$SU(3)$ nonet, $l_q=0$:
      \be
      |F_{\phi P_I}^{\rm asy}(q^2)|=
      |F_{\phi P_I}^{\rm Res}(s_{\rm asy}^{P_I})|
      \left(\frac{s_{\rm asy}^{P_I}}{q^2}\right)^2
      \hspace{15mm} q^2> s_{\rm asy}^{P_I};
      \label{asy-p}
      \en
\item for $M_I=S_I$: $(n_H,n_\lambda,l_q)=(2,1,1)$, there are two
      hadronic fields in the final state (fig.~\ref{phi-eta}) and the hadronic
      helicity flips ($\lambda_{S_I}=0$), but the scalar $q\ov{q}$ state is
      an $n^{2S+1}L_J=1^3P_0$ element of the scalar flavor-$SU(3)$ nonet, hence $l_q=1$:
      \be
      |F_{\phi S_I}^{\rm asy}(q^2)|=
      |F_{\phi S_I}^{\rm Res}(s_{\rm asy}^{S_I})|
      \left(\frac{s_{\rm asy}^{S_I}}{q^2}\right)^\frac{5}{2}
      \hspace{15mm} q^2> s_{\rm asy}^{S_I};
      \label{asy-s}
      \en%
\item finally, for $M_I=S_I^4$: $(n_H,n_\lambda,l_q)=(3,1,1)$, due to the
      tetraquark structure of the scalar meson $S_I^4$, there are three
      hadronic fields in the final state (fig.~\ref{new-hyp-fig}), the hadronic
      helicity flips ($\lambda_{S_I^4}=0$) and, having $S=L=0$, this is
      an $n^{2S+1}L_J=1^1S_0$ element of the scalar flavor-$SU(3)$ diquark nonet, 
      hence $l_q=0$ (note that the quantum numbers do not refer to single quarks,
       but to the diquarks $[qq]$):
      \be
      |F_{\phi S_I^4}^{\rm asy}(q^2)|=
      |F_{\phi S_I^4}^{\rm Res}(s_{\rm asy}^{S_I^4})|
      \left(\frac{s_{\rm asy}^{S_I^4}}{q^2}\right)^3
      \hspace{15mm} q^2> s_{\rm asy}^{S_I^4}.
      \label{asy-s4}
      \en
\ei
\subsection{Analytic continuation in the $q^2$-complex plane}
\label{anal-cont}
As discussed in \textsection\ref{static-ff} a generic
tff $F(q^2)$ is an analytic function in the complex $q^2$ plane with the cut
$(s_{\rm th},\infty)$ (fig.~\ref{cauchy}). This property, together with the vanishing 
asymptotic behavior [eqs.~(\ref{asy-p}-\ref{asy-s4})], allows one to use dispersion 
relations (DR)~\cite{math} to perform the analytic continuation of the tff's. 
\\
By applying the Cauchy theorem~\cite{math} to the integration path $\mathcal{C}$
of fig.~\ref{cauchy}, as the radius $\mathcal{R}$ diverges, we obtain the
DR for the imaginary part:
\be
F(q^2)=\frac{1}{\pi}\int_{s_{\rm th}}^\infty
\frac{\im[F(s)]}{s-q^2}ds,
\label{dr-im}
\en
with $q^2<s_{\rm th}$. Expression~(\ref{dr-im}) states that real values of
the tff $F(q^2)$, below the threshold $s_{\rm th}$, may be computed by 
integrating its imaginary part over the cut.
\\
To exploit directly the experimental data and the asymptotic 
behavior~(\ref{asy-p}-\ref{asy-s4}) for the modulus of the tff's in the 
time-like region, we use the DR for the logarithm\cite{drlog}:
\be
\ln[F(q^2)]=\frac{\sqrt{s_{\rm th}-q^2}}{\pi}\int_{s_{\rm th}}^\infty
\frac{\ln|F(s)|}{(s-q^2)\sqrt{s-s_{\rm th}}}ds\hspace{10mm}
q^2<s_{\rm th},
\label{dr-log}
\en
that connects the real values of the tff below the threshold $s_{\rm th}$ 
to its modulus over the cut. The DR~(\ref{dr-log}) is obtained
by writing down the DR~(\ref{dr-im}) for the function 
$\Psi(z)=\frac{\ln[F(z)]}{\sqrt{s_{\rm th}-z}}$ and it can be used in this form
only if $F(q^2)$ has neither zeros nor poles on the physical sheet. The
parameterizations~(\ref{res-p}-\ref{res-s4}) and (\ref{asy-p}-\ref{asy-s4}) 
in the resonance and asymptotic regions guarantee the absence of singularities
for the tff in physical sheet and allow one to use the DR for the logarithm~(\ref{dr-log}).
\\
Starting from the parameterizations for the modulus of the tff over the cut 
$(s_{\rm th}^{I},\infty)$, by means of the DR~(\ref{dr-log}), we obtained
a parameterization also for the region below the threshold $s_{\rm th}^{I}$ as:
\be
F_{\phi M_I}^{\rm an}(q^2)\!=\!\exp\!\left[\frac{\sqrt{s_{\rm th}^I-q^2}}{\pi}
\!\left(\!
\int_{s_{\rm th}^I}^{s_{\rm asy}^{M_I}}\!\!\!
\frac{\ln|F_{\phi M_I}^{\rm Res}(s)|ds}{(s-q^2)\sqrt{s-s_{\rm th}^I}}+\!\!
\int^\infty_{s_{\rm asy}^{M_I}}\!\!
\frac{\ln|F_{\phi M_I }^{\rm asy}(s)|ds}{(s-q^2)\sqrt{s-s_{\rm th}^I}}\!\right)\!
\right],
\label{par-below}
\en
with $M_I=P_I,\,S_I,\,S_I^4$\,, $I=0,1$\,, $\;q^2<s_{\rm th}^I$ and where, in the 
first integral over the resonance region, we used expressions~(\ref{res-p}-\ref{res-s41}),
while in the second, over the asymptotic region we used~(\ref{asy-p}-\ref{asy-s4}).
\subsection{The overall parameterization}
\label{averall}
The overall parameterization for the tff's, which covers the whole real $q^2$
axis (in principle the whole complex $q^2$ plane), has the threefold 
expression (see fig.~\ref{regions}):
\be
 F_{\phi M_I}(q^2)=\left\{
\begin{array}{lll}
 F_{\phi M_I}^{\rm an}(q^2) &\hspace{5mm} q^2\le s_{\rm th}^I & 
\hspace{5mm}\mbox{[eq.~(\ref{par-below})]}\\
&&\\
 F_{\phi M_I}^{\rm Res}(q^2) &\hspace{5mm} s_{\rm th}^I < q^2\le s_{\rm asy}^{M_I} & 
\hspace{5mm}\mbox{[eqs.~(\ref{res-p})-(\ref{res-s41})]}\\
&&\\
 F_{\phi M_I}^{\rm asy}(q^2) &\hspace{5mm} q^2> s_{\rm asy}^{M_I}  & 
\hspace{5mm}\mbox{[eq.~(\ref{asy-p})-(\ref{asy-s4})]}\\
\end{array}\right. ,
\label{tot-par}
\en
with $M_I=P_I,\,S_I,\,S_I^4$\, and $I=0,1$.
\bfi[ht!]\vspace{-0mm}
\bc
\epsfig{file=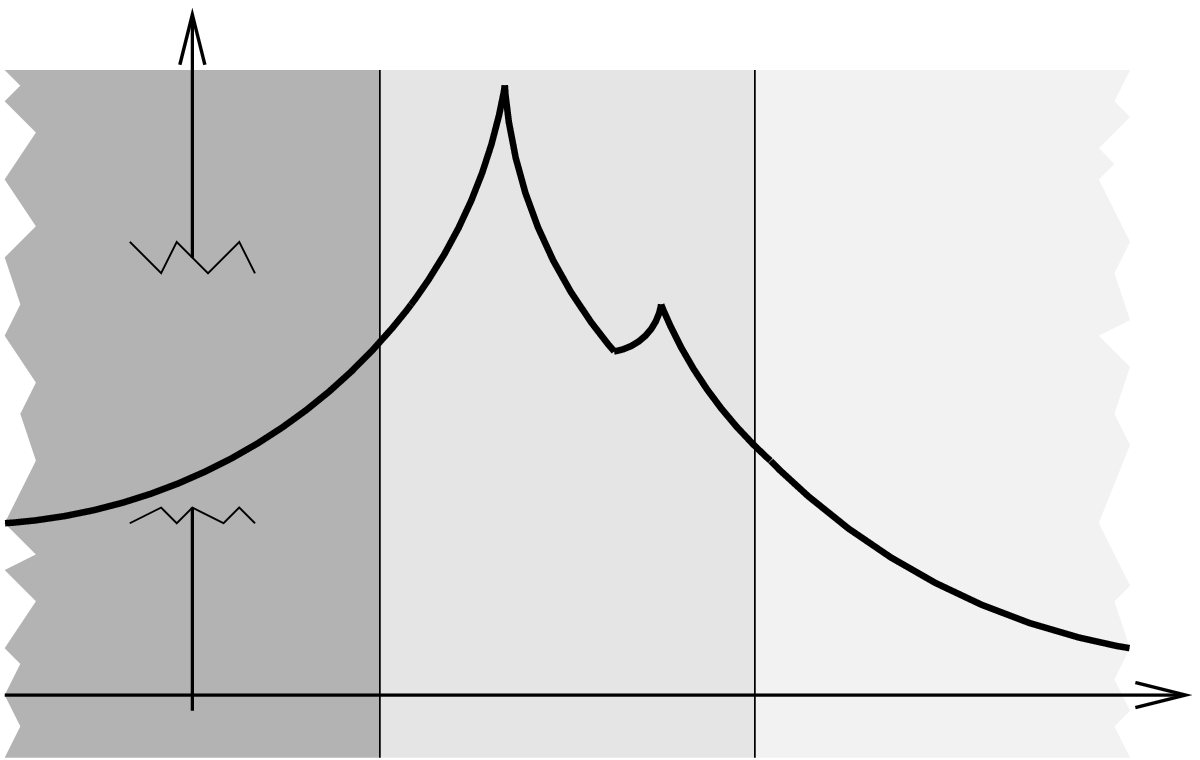,width=100mm,height=60mm}
\put(-234,162){$F_{\phi M_I}(q^2)$}
\put(-262,85){$F_{\phi M_I}^{\rm an}(q^2)$}
\put(-172,75){$F_{\phi M_I}^{\rm Res}(q^2)$}
\put(-82,65){$F_{\phi M_I}^{\rm asy}(q^2)$}
\put(1,13){$q^2$}
\put(-100,5){Asymptotic reg.}
\put(-185,5){Resonance reg.}
\put(-274,5){Analytical reg.}
\caption{\label{regions}
Schematic representation of the three energy regions 
used to parameterize the tff's.
}
\ec
\efi\\
Figure~\ref{regions} shows the regions in which the real $q^2$ axis has
been divided to parametrize the tff's. Starting from the right there are:
\bi
\item the asymptotic region, in light-gray, where the QcR power law
      has been used;
\item the resonance region, in gray, where the tff's are parametrized as a sum
      of resonant contributions;
\item the analytic region, in dark-gray, where the tff's are obtained 
      through the DR's.
\ei
\section{$\chi^2$ definition}
\label{chi2-def}
\subsection{Free parameters and fixed values}
\label{free-par}
The free parameters of the expression~(\ref{tot-par}) are all located in
the resonant part of the definition $F_{\phi M_I}^{\rm Res}(q^2)$. Different choices
may be done about masses and widths of the contributing vector resonances.
For the leading contributions, i.e.: the $\phi(1020)$ in case of $M_0=P_0,S_0$, 
the $\omega(782)$ in case of $M_0=S_0^4$ and the $\rho^0(770)$ in case of 
$M_1=P_1,S_1,S_1^4$, fixed values of masses and widths
may be used, because many of these quantities are known with a high degree of accuracy. 
Other masses and widths, namely those of $\phi$-, $\omega$- and $\rho$-recurrences, 
being poorly known, are left free, indeed one of the aims of
the analysis
is a better determination of their values.
\\
Concerning the couplings, while some values of $F_V$ may be extracted from data 
on decay rates $\Gamma(V\ra e^+e^-)$ and then they are assumed as fixed, 
all the other couplings $F_V$ and $g^V_{\phi M_I}$, and the relative phases 
$\Phi_{VM_I}$ are considered as free parameters.
\\
Finally, also the threshold energies $s_{\rm asy}^{M_I}$, from which the asymptotic behavior
is assumed, are left free.
\subsection{Constraints}
\label{condizioni}
Two kinds of conditions may be used to fix the free parameters of
the procedure: the experimental and theoretical constraints.\\
In general, the experimental constraints come from data for three
different processes:
\bi
\item the radiative decay $\phi\ra M_I\gamma$ gives the real value of the tff
      at $q^2=0$;
\item the conversion decay  $\phi\ra M_I e^+e^-$ covers the interval
      $(2m_e)^2\le q^2\le (M_\phi-M_{M_I})^2$, it gives the real value of
      the tff below $s_{\rm th}^I$ and its modulus above this threshold;
\item the annihilation $e^+e^-\ra \phi M_I$ provides a measurement of the 
      modulus of the tff above the physical threshold, i.e. 
      for $q^2\ge(M_\phi+M_{M_I})^2$.
\ei
The theoretical condition requires analyticity for the tff, imposing the
continuity of its first derivative across the threshold $s_{\rm th}^I$,
by means of the DR~(\ref{dr-log}), i.e., it forces the identity:
\be
\left.\frac{1}{F_{\phi M_I}^{\rm Res}}\frac{dF_{\phi M_I}^{\rm Res}}{ds}
\right|_{s_{\rm th}^I}\!\!\!\!
=\lim_{\epsilon\ra 0^+}\frac{1}{2\sqrt{\epsilon}}
\int_{s_{\rm th}^I}^\infty\!\!\frac{(s_{\rm th}^I- s+ \epsilon)
\ln|F_{\phi M_I}(s)|}{\sqrt{s- s_{\rm th}^I}(s- s_{\rm th}^I+\epsilon)^2}ds.
\label{der1}
\en
This condition represents a further constraint which concerns both
the resonance and the asymptotic part of the parameterization~(\ref{tot-par}).
Note that the DR for the logarithm~(\ref{dr-log}) automatically ensures
continuity for the tff (zero derivative) across $s_{\rm th}^I$.
\subsection{$\chi^2$ expression}
\label{chi2-expression}
Following the previous discussion we define a $\chi^2$ with two
contributions:
\be
\chi^2_{M_I}=\chi^2_{M_I,\rm exp}+\chi^2_{M_I,\rm theory}.
\en
The experimental contribution is defined as:
\be
\chi_{M_I,\rm exp}^2=\sum_{j}\left[
\frac{|F_{\phi M_I}(s^{M_I}_j)|-F_{j}^{M_I}}
{\delta F_{j}^{M_I}}\right]^2,
\label{chi2}
\en
where $\{s^{M_I}_j,F_{j}^{M_I}\pm\delta F_{j}^{M_I}\}$ is the total 
set of data on the $\phi M_I$ tff, as we have discussed in the previous sections,
these data may be extracted from decay rates and cross sections of different 
processes which involve the mesons $\phi$ and $M_I$.
\\
The theoretical contribution, which forces the continuity of the first
derivative of the tff across the threshold $s_{\rm th}^I$, has the form
[eq.~(\ref{der1})]:
\be
\chi_{M_I,\rm theory}^2=\tau\cdot
\left[\left.\frac{1}{F_{\phi M_I}^{\rm Res}}\frac{dF_{\phi M_I}^{\rm Res}}{ds}
\right|_{s_{\rm th}^I}\!\!\!\!
-\lim_{\epsilon\ra 0^+}\frac{1}{2\sqrt{\epsilon}}
\int_{s_{\rm th}^I}^\infty\!\!\frac{(s_{\rm th}^I- s+ \epsilon)
\ln|F_{\phi M_I}(s)|}{\sqrt{s- s_{\rm th}^I}(s- s_{\rm th}^I+\epsilon)^2}ds
\right]^2\!\!,
\en
where $\tau$ is a free parameter. Since the theoretical condition has to
be exactly verified in order to satisfy the analyticity requirement, the values of
$\tau$ can be chosen large enough to guarantee the vanishing of $\chi_{M_I,\rm theory}^2$.
The theoretical condition looks like a super-convergence relation, 
indeed it constrains the value of an integral of the tff over the cut and
it represents a crucial constraint in the resonance region, below the physical
threshold, where there are no data.
\section{Possible aims}
\label{aims}
\subsection{Looking for masses, widths and coupling constants}
The main aim of this analysis procedure is to provide a practical tool,
based on theory and data, to find masses, widths and coupling constants of 
vector mesons, like $\phi$-, $\omega$- and $\rho$-recurrences, that are still 
poorly known.
The key points of this method, which make it a step forward with
respect to the usual fit procedures, are:
\bi
\item the analysis is performed in terms of tff's instead of single
      observables like: decay rates and cross sections, this allows 
      to use at the same time, as measurements of the same quantity
      in different energy regions, data coming from different processes;
\item by using an analytic continuation procedure based on DR's, additional
      information about the quark structure of the mesons, coming from the 
      QcR power law used for the asymptotic behavior, may be 
      included in the computation;
\item thanks to overall validity of the parameterization, we may analyze
      resonances lying in energy regions not experimentally accessible,
      like the unphysical region $[(M_\phi-M_{M_I})^2,(M_\phi+M_{M_I})^2]$,
      or at energies hardly reachable, e.g. near thresholds.
\ei
In other words, the constraints from data and theory shape, through 
the analyticity requirement, the tff's at every value of $q^2$.
\subsection{Investigating quark structures}
Since all the three steps of this procedure: parameterization in the resonance
region (\textsection\ref{select}), definition of the asymptotic behavior
(\textsection\ref{asy-beha}) and extension below the threshold $s_{\rm th}^I$
(\textsection\ref{anal-cont}), depend on the assumed quark structure for the 
meson $M_I$, we may exploit this method to test the agreement with data of 
different possible structures and then to gain information on the nature of the
meson $M_I$.
\\
The most interesting cases are those of scalar mesons $S_I$ and $S_I^4$.
In fact we noted that by passing from the $q\ov{q}$ to the $[qq][\ov{qq}]$ 
structure, the tff parameterization changes drastically in all the three
regions. 
\\
In particular for $I=0$, the tetraquark hypothesis for the scalar meson
implies additional $\omega$-family contributions in the resonance region and a slightly
faster vanishing asymptotic behavior, the power goes from $5/2$ to $3$.
\\
In case of $I=1$ the situation appears more drastic, in fact while the
conversion $\phi\gamma S_1$, with a $q\ov{q}$-scalar meson is 
OZI-suppressed (fig.~\ref{rho-pi0}), the same conversion $\phi\gamma S_1^4$,
but with a tetraquark $S_1^4$, should be, instead, OZI-allowed (fig.~\ref{s4-i1}).
\section{Conclusion}
\label{conclu}
We have defined a general procedure which, combining: the description
of the tff's in terms of vector meson propagators, the QcR asymptotic power 
law, and the analyticity requirement by means of DR's, provides a parameterization
for the tff's valid in the whole $q^2$-real axis. 
The experimental and theoretical constraints, that individually
concern only certain energy regions, have been ``propagated'' to all
values of $q^2$.
\\
The application of this procedure may be twofold. 
\\
By assuming as known the quark structure of the mesons 
under consideration, we can look for vector resonances, i.e. excited states 
of the $\phi(1020)$, $\omega(782)$ and $\rho^0(770)$, which couple with the 
studied final state. 
\\
By exploiting the fact that this procedure is strongly dependent on 
the quark structure of the final mesons, 
we may test how well different 
hypotheses about the this  structure agree with data.
\acknowledgments{
I warmly thank Rinaldo Baldini, Gino Isidori, Lia Pancheri, Mike Sokoloff, 
and Adriano Zallo for precious 
and instructive discussions on the subject of this work.
}
\section*{}
%
%

%
%

\begin{thebibliography}{99}
%
\bibitem{flavor} B.~Di Micco [KLOE Collaboration], arXiv:hep-ex/0410072;\\
%
V.~Druzhinin  [BABAR Collaboration],  arXiv:hep-ex/0601020.
%
\bibitem{jlab} O.~Gayou {\it et al.}  [Jefferson Lab Hall A Collaboration],
  Phys.\ Rev.\ Lett.\  {\bf 88} (2002) 092301
  [arXiv:nucl-ex/0111010].
%
\bibitem{landsberg}L.~G.~Landsberg, Phys.\ Rept.\  {\bf 128} (1985) 301.
%
\bibitem{brodsky} V.~A.~Matveev, R.~M.~Muradian and A.~N.~Tavkhelidze,
  Lett.\ Nuovo Cim.\  {\bf 7}, 719 (1973);\\
S.~J. Brodsky, G.~R. Farrar, Phys. Rev. D {\bf 11} (1975) 1309;\\
  S.~J.~Brodsky and B.~T.~Chertok, Phys.\ Rev.\ D {\bf 14} (1976) 3003.
%
\bibitem{ozi} G.~Zweig, CERN report S419/TH412 (1964), unpublished;\\ 
S.~Okubo, Phys. Lett, {\bf 5} (1963) 165;\\
I.~Iizuka, K.~Okuda, O.~Shito, Prog. Theor. Phys. {\bf 35} (1966) 1061.
%
\bibitem{4q}R.~L.~Jaffe, Phys.\ Rept.\  {\bf 409} (2005) 1
  [Nucl.\ Phys.\ Proc.\ Suppl.\  {\bf 142} (2005) 343]
  [arXiv:hep-ph/0409065] and rferrences therein.
%
\bibitem{hely}S.~J. Brodsky, G.~P. Lepage, Phys. Rev. D {\bf 24} (1981) 2848.
S.~J.~Brodsky and G.~F.~de Teramond, Phys.\ Lett.\ B {\bf 582} (2004) 211
  [arXiv:hep-th/0310227].
%
\bibitem{math} See for instance: E.C. Tichmarsch, {\it The theory of functions}, 
  London, Oxford University Press, 1939.
%
\bibitem{drlog}B.~V.~Geshkenbein, Yad.\ Fiz.\  {\bf 9} (1969) 1232.
%
\end{thebibliography}
\end{document}